\newif\ifrevtexavailable
\IfFileExists{revtex4.cls}{\revtexavailabletrue}{\revtexavailablefalse}
\ifrevtexavailable
\documentclass[showpacs,amsmath,amsfonts,amssymb,aps,superscriptaddress]{revtex4}
\else
\documentclass[10pt,a4paper]{article}
\usepackage[textwidth=508pt,left=54pt,top=54pt,bottom=54pt]{geometry}
\usepackage[numbers,sort&compress]{natbib}
\usepackage{authblk}
\newcommand{\pacs}[1]{}
\fi
\usepackage[T1]{fontenc}
\usepackage[utf8]{inputenc}
\usepackage{lmodern}
\ifrevtexavailable\fi

\usepackage{tabularx}
\usepackage{xcolor}
\definecolor{RED}{rgb}{1,0,0}
\usepackage{booktabs}
\usepackage{url}
\newcommand{\rev}[1]{{\color{red}#1}}

\usepackage{comment}
\usepackage{tikz}
\usetikzlibrary{decorations.pathreplacing}
\usepackage{enumitem}
\usepackage{graphicx}% Include figure files
\usepackage[font=scriptsize]{caption} % Small-sized captions
\usepackage{float}
\usepackage{dcolumn}% Align table columns on decimal point
\usepackage{bm}% bold math
\usepackage{mathrsfs}
\usepackage{amsmath}

\usepackage{amsfonts}

\begin{document}

\title{Quasinormal modes of Hayward black holes from Schwarzschild to extremality:\\ a comparative spectral analysis}% Force line breaks with \\
\ifrevtexavailable
\author{Davide Batic}
\email{davide.batic@ku.ac.ae}
\affiliation{Mathematics Department, Khalifa University of Science and Technology, PO Box 127788, Abu Dhabi, United Arab Emirates}

\author{Denys Dutykh}
\email{denys.dutykh@ku.ac.ae}
\affiliation{Mathematics Department, Khalifa University of Science and Technology, PO Box 127788, Abu Dhabi, United Arab Emirates}

\else
\author[1]{Davide Batic}
\author[1]{Denys Dutykh}
\affil[1]{Mathematics Department, Khalifa University of Science and Technology, PO Box 127788, Abu Dhabi, United Arab Emirates}
\affil[]{\texttt{davide.batic@ku.ac.ae}; \texttt{denys.dutykh@ku.ac.ae}}
\fi

\date{\today}

\date{\today}% It is always \today, today,

\ifrevtexavailable\else\maketitle\fi
\begin{abstract}
We study scalar and electromagnetic quasinormal modes of the Hayward black hole and the axial spectrum of a specified Regge--Wheeler-type effective model, from the Schwarzschild limit to exact extremality. Multiprecision Chebyshev collocation reproduces most available low-lying oscillatory benchmarks and supplies further overtone candidates. The spin-2 equation is model dependent. It differs from the effective-source potential used in recent Hayward studies, and we quantify representative frequency differences. The discretisation also produces purely imaginary eigenvalue sequences, whose physical classification requires special care because highly reproducible imaginary-axis ladders need not be poles of the continued Green function. Accordingly, the imaginary-axis entries reported here are retained as unclassified spectral candidates, not established QNMs. We quantify their finite-table gap dispersion and sensitivity to the number of included gaps. Their spacings near one quarter in $M\omega$, and the extremal product near $1/3$ with the throat radius, remain properties of the tabulated candidates, and they do not establish universal QNM spacing or throat-controlled damping.
\end{abstract}

\pacs{XXX}% PACS, the Physics and Astronomy
\ifrevtexavailable\maketitle\fi

\section{Introduction}
\label{Intro}

Removing the central curvature singularity is only the first test of a proposed regular black-hole geometry. A credible model must also possess a controlled dynamical response and must reduce to the classical theory in the regime where general relativity is well tested. Quasinormal modes (QNMs) provide one of the sharpest probes of that response. They are the characteristic complex resonances of an open black-hole system and determine the oscillatory decay of linear perturbations. QNMs therefore connect mathematical questions of stability and scattering with the ringdown of compact objects observed through gravitational waves \cite{Kokkotas1999LR, Berti2009CQG, Konoplya2011RMP, Abbott2016PRL}. For test fields on a fixed background, the spectrum is determined by the geometry, the spin of the field, and the boundary conditions. For genuine gravitational perturbations of an effective or non-vacuum geometry, the perturbation dynamics and the treatment of the effective source are also part of the physical specification.

The fundamental mode is usually the most robust component of a ringdown signal, but it is not the whole spectrum. Numerical-relativity studies have shown that including overtones can extend a QNM description towards the peak of a merger waveform \cite{Giesler2019PRX, Giesler2025PRD}. At the same time, the identification and interpretation of individual subdominant tones can depend sensitively on the fitting model, the chosen start time, and non-QNM contributions \cite{Nee2023PRD}. This observational subtlety does not diminish the theoretical value of the higher spectrum. Overtones and strongly damped modes sample the radial problem differently from the fundamental mode and can respond much more strongly to changes in the near-horizon and interior analytic structure. A numerical method intended to characterise a quantum-corrected or regular black hole should therefore be tested not only on the lowest resonance, but also on extended oscillatory sequences and on branches close to the imaginary-frequency axis.

The Hayward spacetime is a particularly useful setting in which to pursue this programme. It was introduced as a simple nonsingular model of black-hole formation and evaporation, with an asymptotically Schwarzschild exterior and a regular de Sitter-like core replacing the central singularity \cite{Hayward2006PRL}. Depending on its deformation parameter, the static geometry possesses two distinct horizons, a degenerate horizon at extremality, or no horizon. Its simplicity has made it a standard representative of the broader class of regular black holes. A closely related functional form also arises in renormalisation-group-improved constructions inspired by asymptotically safe gravity (ASG), once a particular identification between the running scale and a curvature scale is adopted \cite{Held2019JCAP}. This connection is physically suggestive but should not be overstated since the Hayward metric is not a unique prediction of ASG, and different scale-setting prescriptions or microscopic completions can lead to different effective geometries and different perturbation equations. In the present work, it is treated as a specified effective background whose spectral consequences can be calculated and compared in a controlled manner.

The QNM spectrum of the Hayward geometry has already been examined from several complementary perspectives. Konoplya et al. combined time-domain evolution, WKB approximations, and a frequency-domain method to study scalar and electromagnetic perturbations, including selected overtones and modes close to the imaginary axis \cite{Konoplya2022JCAP}. Malik subsequently derived inverse-multipole approximations for scalar, Dirac, and Maxwell fields and compared them with WKB and time-domain estimates \cite{Zainab2024EPL}. More recently, axial gravitational perturbations have been considered using effective-source closures. For instance, Malik investigated QNMs in connection with grey-body factors \cite{Malik2025IJTP}, while Bolokhov and Skvortsova computed the fundamental mode and first overtone with WKB--Pad\'e and time-domain methods and found that the deformation generally raises the oscillation frequency and lowers the damping rate \cite{Bolokhov2026EPJC}. These works establish important low-lying benchmarks and demonstrate the sensitivity of subdominant modes. They do not, however, provide a single high-precision map of the scalar, electromagnetic, and axial spin-2 spectra from the Schwarzschild limit to the exactly extremal geometry, including extended overtone families and the deeply damped purely imaginary sector.

This remaining gap is partly methodological. A converged finite spectral eigenvalue need not be an isolated pole of the analytically continued Green function. The Schwarzschild response contains a branch-cut contribution on the negative imaginary axis (NIA) \cite{Leaver1986PRD}. Spectral eigenfunction and boundary-sector checks are therefore essential \cite{FortunaVega2023EPJC}. Our subsequent study of axial Schwarzschild perturbations \cite{BaticDutykhSukaiti2026NIA} makes the distinction explicit because a highly reproducible 68-point imaginary-axis ladder fails an independent lateral Jost/Wronskian pole test. This result changes the interpretation of the imaginary-axis sequences considered here. It is not a numerical classification of the nonzero-deformation or extremal Hayward problem. The purpose of this paper is to document the oscillatory spectra and the finite-discretisation candidates obtained in the Hayward part of the comparative programme \cite{Batic2026EPJC,Batic2026PRD}, while separating their evidentiary status. The non-extremal and exactly extremal radial equations are formulated separately, because the horizon changes from a simple to a double zero. Increasing-order checks and multiprecision arithmetic provide useful tests of the finite pencils. They do not, by themselves, select the physical analytic continuation on the NIA. We retain the tabulated imaginary-axis values for reproducibility, but withhold their identification as QNMs. An additional qualification concerns spin 2. The Hayward line element does not specify how its effective source is perturbed. Here, the axial model is defined by $U_2=F[\ell(\ell+1)/r^2-3F'/r]$, and it is not derived from a specified coupled metric--matter action. Notice that the effective-source closure used in \cite{Malik2025IJTP,Bolokhov2026EPJC} gives a different potential. We display their difference and compare published frequencies at matching deformation and multipole values. The resulting model dependence should be distinguished from a numerical-method error. The supported low-lying results show modest increases of representative oscillation frequencies and reduced damping as the Hayward deformation grows. The largest relative change among the fundamental modes considered occurs in the adopted axial quadrupole. In the imaginary-axis tables, gaps are often close to $0.25$ in $M\omega$, but some sequences contain large gaps or near-zero roots. We report explicit finite-window statistics and their limitations. A common numerical formulation across geometries can generate common representation effects, so similarities to \cite{Batic2026EPJC,Batic2026PRD} are not independent evidence for physical QNM universality.

The manuscript is organized as follows. Section~II gives the geometry, the effective potentials, and the formal asymptotic factorisations. Section~III distinguishes finite-pencil convergence from physical pole identification and states which validations are available. Section~IV presents the spectra, compares axial closures, and analyses the tabulated imaginary-axis gaps. Section~V revisits the comparison across the trilogy. Section~VI states the conclusions and outstanding tests.

\section{Prolegomena on the Hayward metric}

The Hayward line element in Boyer–Lindquist coordinates is (see \cite{Konoplya2022JCAP, Zainab2024EPL})  
\begin{equation}\label{LE}
ds^2=-F(r)dt^2+\frac{dr^2}{F(r)}+r^2 d\vartheta^2+r^2\sin^2{\vartheta}d\varphi^2,
\quad
F(r)=1-\frac{2r^2 M^{-2}}{r^3 M^{-3}+\gamma}.
\end{equation}  
The standard Schwarzschild metric is recovered in the limit $\gamma = 0$. For $\gamma > 0$, the central singularity at $r = 0$ is resolved, yielding a regular spacetime. In the parameter range $0 < \gamma < 32/27$, the metric describes a non-extremal regular black hole featuring two distinct horizons, namely the event horizon $r_h$ and the Cauchy horizon $r_c$. At the critical value $\gamma = 32/27$, these horizons coincide at $r_e = 4M/3$, which corresponds to an extremal Hayward black hole. For $\gamma > 32/27$, no horizons are present, and the solution represents a compact self-gravitating, horizonless object often interpreted as a gravitational vacuum condensate or droplet. If we introduce the rescaling $\rho=r/M$ and we impose $F(\rho_h)=0$, it is not difficult to verify that the location of the event horizon is obtained by solving the equation
\begin{equation}
\rho_h^3-2\rho_h^2+\gamma=0.    
\end{equation}
As shown in Table~\ref{table:event}, increasing the parameter $\gamma$ up to its critical value causes $\rho_h$ to approach the event horizon of the extremal Hayward black hole.
\begin{table}[ht]
\small\setlength{\tabcolsep}{4.5pt}
%\centering
\caption{Representative numerical values of the rescaled event horizon radius $\rho_h$ for various values of the parameter $\gamma$ considered by \cite{Konoplya2022JCAP, Zainab2024EPL}.}
\label{table:event}
\vspace*{1em}
\begin{tabular}{||c|c|c|c||}
\hline\hline
$\gamma$          & $\rho_h$        & $\gamma$  & $\rho_h$\\ [0.5ex]
\hline\hline
$0$               & $2$             & $0.50$    & $1.854637680$\\
$0.02$            & $1.994974779$   & $16/27$   & $1.821367205$\\
$0.04$            & $1.989898212$   & $0.70$    & $1.778760216$\\
$0.06$            & $1.984768897$   & $20/27$   & $1.761189018$\\
$0.08$            & $1.979585370$   & $24/27$   & $1.688059257$\\
$0.10$            & $1.974346098$   & $0.90$    & $1.681807499$\\
$0.12$            & $1.969049477$   & $1.0$     & $1.618033989$\\
$4/27$            & $1.961494567$   & $28/27$   & $1.589578102$\\
$8/27$            & $1.919590161$   & $1.10$    & $1.530246862$\\
$0.30$            & $1.918491788$   & $1.18$    & $1.383622516$\\
$12/27$           & $1.873358620$   & $32/27$   & $4/3$\\
[1ex]
\hline\hline 
\end{tabular}
\end{table}

For a massless scalar test field on the spherically symmetric metric \eqref{LE}, separation with time dependence $e^{-i\omega t}$ gives a radial wave equation \cite{Batic2019EPJC}. We write it together with the electromagnetic test-field equation and the adopted axial effective model as

\begin{equation}\label{ODE01}
    F(r)\frac{d}{dr}\left(F(r)\frac{d\psi_{\omega\ell\epsilon}}{dr}\right)+\left[\omega^2-U_\epsilon(r)\right]\psi_{\omega\ell\epsilon}(r)=0,\qquad
    U_\epsilon(r)=F(r)\left[\frac{\epsilon}{r}\frac{dF}{dr}+\frac{\ell(\ell+1)}{r^2}\right],\qquad
    \epsilon=1-s^2
\end{equation}

where $(s,\epsilon)=(0,1),(1,0),(2,-3)$ and the radiative multipoles satisfy $\ell\geq s$. For $s=2$, this is a definition of the effective axial model, not a consequence of the Klein--Gordon equation. Writing $F=1-2m(r)/r$, the adopted potential and the effective-source potential of \cite{Malik2025IJTP,Bolokhov2026EPJC} are
\begin{align}
U_2^{\rm here}&=F\left[\frac{\ell(\ell+1)}{r^2}-\frac{6m}{r^3}+\frac{6m'}{r^2}\right],\label{eq:axial_here}\\
U_2^{\rm src}&=F\left[\frac{\ell(\ell+1)}{r^2}+\frac{2(F-1)}{r^2}-\frac{F'}{r}\right]
=F\left[\frac{\ell(\ell+1)}{r^2}-\frac{6m}{r^3}+\frac{2m'}{r^2}\right].\label{eq:axial_source}
\end{align}
For $m(r)=Mr^3/(r^3+\gamma M^3)$, their exact difference is
\begin{equation}\label{eq:axial_difference}
U_2^{\rm here}-U_2^{\rm src}=\frac{4Fm'}{r^2}
=\frac{12\gamma M^4 F}{(r^3+\gamma M^3)^2}.
\end{equation}
It vanishes in the Schwarzschild limit and is positive outside the event horizon for $\gamma>0$. The two problems therefore have the same classical limit but different deformed spectra. A potential ordering alone does not prove an ordering of complex QNM frequencies. All axial tables below use $U_2^{\rm here}$ while  Sec.~\ref{sec:axial_results} quantifies representative differences from the published source-closure frequencies.

By means of the substitution $x=\rho/\rho_h$, the above equation can be recast in the equivalent form
\begin{equation}\label{ourODE}
F(x)\frac{d}{dx}\left(F(x)\frac{d\psi_{\Omega\ell\epsilon}}{dx}\right)+\left[\rho_h^2\Omega^2-V_\epsilon(x)\right]\psi_{\Omega\ell\epsilon}(x)=0,\quad
F(x)=1-\frac{2\rho_h^2 x^2}{\rho_h^3 x^3 +\gamma},\quad
\Omega=M\omega
\end{equation} 
with effective potential given by 
\begin{equation}\label{Veff}
V_\epsilon(x)=F(x)\left[\frac{\epsilon}{x}\frac{dF}{dx}+\frac{\ell(\ell+1)}{x^2}\right].
\end{equation}
The following analysis computes oscillatory QNM benchmarks and additional spectral candidates for the problem stated in \eqref{ourODE}. For this purpose, we represent $\omega$ as $\omega = \omega_R + i\omega_I$, where $\omega_I < 0$ ensures that the perturbation is damped in time. The boundary conditions are set so that the radial field exhibits inward radiation at the event horizon and outward radiation at spatial infinity. This necessitates a thorough examination of the solution's asymptotic behaviour in \eqref{ourODE}, both near the event horizon ($x \to 1^{+}$) and at large spatial distances ($x \to +\infty$). Moreover, to compute the QNMs using the spectral method, we must recast the differential equation \eqref{ourODE} and the appropriate boundary conditions over the compact interval $[-1,1]$. This adjustment is necessary as the method expands the regular part of the eigenfunctions by means of Chebyshev polynomials. 

\subsection{The non-extreme case}

This scenario focuses on the parameter $\gamma$ satisfying the inequality $0\leq\gamma<32/27$. We begin by observing that the event horizon corresponds to a simple zero of the function $F(x)$ defined in \eqref{ourODE}. Furthermore, imposing the condition $F(1) = 0$, which is equivalent to $\gamma=2\rho_h^2-\rho_h^3$,  enables us to recast $F(x)$ in the form
\begin{equation}
F(x)=\frac{[\rho_h^2 x^2+(\rho_h-2)(x+1)](x-1)}{\rho_h x^3+2-\rho_h}.
\end{equation}
To establish the QNM boundary conditions at the event horizon and at infinity, we first need to determine the asymptotic behaviour of the radial solution $\psi_{\Omega\ell\epsilon}$ as $x \to 1^{+}$ and as $x \to +\infty$. We can then extract the QNM boundary conditions from this asymptotic data. Concerning the asymptotic behaviour as $x\to 1^+$, it is convenient to reformulate \eqref{ourODE} in the form
\begin{equation}\label{ODEZ}
\frac{d^2\psi_{\Omega\ell\epsilon}}{dx^2}+p(x)\frac{d\psi_{\Omega\ell\epsilon}}{dx}+q(x)\psi_{\Omega\ell\epsilon}(x)=0,\quad
p(x)=\frac{F^{'}(x)}{F(x)},\quad
q(x)=\frac{\rho_h^2\Omega^2-V_\epsilon(x)}{F^2(x)}.   
\end{equation}
Since $p$ and $q$ have poles of order one and two at $x = 1$, respectively, this point is classified as a regular singular point of \eqref{ODEZ}, according to Frobenius theory \cite{Ince1956}. Hence, we can construct solutions of the form
\begin{equation}
\psi_{\Omega\ell\epsilon}(x)=(x-1)^\alpha\sum_{\kappa=0}^\infty a_\kappa(x-1)^\kappa.
\end{equation}
The leading behavior at $x=1$ is represented by the term $(x-1)^\alpha$ where $\alpha$ is determined by the indicial equation
\begin{equation}\label{indicial}
\alpha(\alpha-1)+P_0\alpha+Q_0=0
\end{equation}
with
\begin{equation}
P_0=\lim_{x\to 1}(x-1)p(x)=1,\qquad
Q_0=\lim_{x\to 1}(x-1)^2 q(x)=\left(\frac{2\rho_h\Omega}{3\rho_h-4}\right)^2.
\end{equation}
The roots of \eqref{indicial} are $\rho_\pm = \pm 2i \rho_h\Omega/(4-3\rho_h)$ and the correct QNM boundary condition at $x=1$ reads
\begin{equation}\label{QNMBCz1}
\psi_{\Omega\ell\epsilon}\underset{{x\to 1^+}}{\longrightarrow} (x-1)^{-i \rho_h a\Omega},\quad a=\frac{2}{3\rho_h-4}.
\end{equation}
From the above expression, we immediately see that $a$ is always positive since $\rho_h>4/3$. Regarding the asymptotic behaviour as $x\to+\infty$, we start by observing that
\begin{equation}
p(x) = \sum_{\kappa=0}^\infty\rev{\frac{\mathfrak{f}_\kappa}{x^\kappa}} = \mathcal{O}\left(\frac{1}{x^2}\right), \qquad
q(x) = \sum_{\kappa=0}^\infty\rev{\frac{\mathfrak{g}_\kappa}{x^\kappa}}=\rho_h^2\Omega^2+\frac{4\rho_h\Omega^2}{x}+\mathcal{O}\left(\frac{1}{x^2}\right).
\end{equation}
Consequently, the asymptotic behaviour of the solutions to equation \eqref{ODEZ} can be deduced using the method outlined in \cite{Olver1994MAA}. Given that at least one of the coefficients $\mathfrak{f}_0$, $\mathfrak{g}_0$, $\mathfrak{g}_1$ is nonzero, a formal solution to \eqref{ODEZ} is represented by \cite{Olver1994MAA}
\begin{equation}\label{olvers}
\psi^{(j)}_{\Omega\ell\epsilon}(x) = x^{\mu_j}e^{\lambda_j x}\sum_{\kappa=0}^\infty\frac{a_{\kappa,j}}{x^\kappa}, \qquad j \in \{1,2\},
\end{equation}
where $\lambda_1$, $\lambda_2$, $\mu_1$ and $\mu_2$ are the roots of the characteristic equations
\begin{equation}\label{chareqns}
   \lambda^2+\mathfrak{f}_0\lambda+\mathfrak{g}_0=0,\quad
   \mu_j=-\frac{\mathfrak{f}_1\lambda_j+\mathfrak{g}_1}{\mathfrak{f}_0+2\lambda_j}.
\end{equation}
A straightforward computation shows that $\lambda_\pm = \pm i\rho_h\Omega$ and $\mu_\pm = \pm 2i\Omega$. As a result, the QNM boundary condition at space-like infinity can be expressed as
\begin{equation}\label{QNMBCzinf}
    \psi_{\Omega\ell\epsilon}\underset{{x\to +\infty}}{\longrightarrow} x^{2i\Omega}e^{i\rho_h\Omega x}.
\end{equation}
We extract the formal ingoing and outgoing asymptotic factors by the following transformation
\begin{equation}\label{Ansatz}
\psi_{\Omega\ell\epsilon}(x) = x^{i(2+a\rho_h)\Omega}(x-1)^{-ia\rho_h\Omega}e^{i\rho_h\Omega(x-1)} \Phi_{\Omega\ell\epsilon}(x).
\end{equation}
Notice that regularity of the remainder is a numerical endpoint condition, and on the NIA, it is not equivalent to selecting the physical radiation sectors, as discussed in Sec.~\ref{sec:pole_classification}.  If we substitute \eqref{Ansatz} into \eqref{ourODE}, we end up with the following ordinary differential equation for the radial eigenfunctions, namely
\begin{equation}\label{ODEznone}
    P_2(x)\Phi^{''}_{\Omega\ell\epsilon}(x) + P_1(x)\Phi^{'}_{\Omega\ell\epsilon}(x) + P_0(x)\Phi_{\Omega\ell\epsilon}(x) = 0
\end{equation}
with
\begin{eqnarray}
P_2(x)&=&x^2(x-1)^2 F^2(x),\\
P_1(x)&=&x(x-1)F(x)\left\{x(x-1)F^{'}(x)+i\Omega F(x)\left[2\rho_h x^2+2x(2-\rho_h)-2a\rho_h-4\right]\right\},\\
P_0(x)&=&-\Omega^2 Q_+(x)Q_{-}(x)+i\Omega F(x)L(x)-x^2(x-1)^2 V_\epsilon(x),\\
Q_\pm(x)&=&F(x)[(x-1)(\rho_h x+2)-a\rho_h]\pm \rho_h x(x-1),\\
L(x)&=&x(x-1)[\rho_h x^2+(2-\rho_h)x-2-a\rho_h]F^{'}(x)-F(x)[2x^2-(2x-1)(2+a\rho_h)].
\end{eqnarray}
Let us now introduce the transformation $x=2/(1-y)$ mapping the point at infinity and the event horizon to $y = 1$ and $y = -1$, respectively. Furthermore, a dot denotes differentiation with respect to the new variable $y$. Then, equation \eqref{ODEznone} becomes
\begin{equation}\label{ODEynone}
    S_2(y)\ddot{\Phi}_{\Omega\ell\epsilon}(y) + S_1(y)\dot{\Phi}_{\Omega\ell\epsilon}(y) + S_0(y)\Phi_{\Omega\ell\epsilon}(y) = 0,
\end{equation}
where
\begin{eqnarray}
  S_2(y) &=&(1+y)^2 F^2(y), \label{S2onone} \\
  S_1(y) &=& i\Omega\frac{1+y}{(1-y)^2}F^2(y)\left[8\rho_h+4(2-\rho_h)(1-y)-2(2+a\rho_h)(1-y)^2\right]\nonumber\\
  &&-2\frac{(1+y)^2}{1-y}F^2(y)+(1+y)^2 F(y)\dot{F}(y), \label{S1onone}\\
  S_0(y) &=& \Omega^2\Sigma_2(y)+i\Omega\Sigma_1(y)+\Sigma_0(y) \label{S0onone}
\end{eqnarray}
with
\begin{eqnarray}
\Sigma_2(y) &=& \frac{4\rho_h^2(1+y)^2}{(1-y)^4}-\frac{F^2(y)}{(1-y)^4}\left\{2(1-y^2)-\rho_h[ay^2-2y(1+a)+a-2]\right\}^2,\\
\Sigma_1(y) &=&F(y)\left\{
\frac{1+y}{(1-y)^2}[4\rho_h+2(2-\rho_h)(1-y)-(2+a\rho_h)(1-y)^2]\dot{F}(y)\right.\nonumber\\
&&-\left.\frac{F(y)}{(1-y)^2}[8-(2+a\rho_h)(3+y)(1-y)]
\right\},\\
\Sigma_0(y)&=&-\frac{4(1+y)^2}{(1-y)^4}V_\epsilon(y).
\end{eqnarray}
Notice that we must also require that $\Phi_{\Omega\ell\epsilon}(y)$ is regular at $y=\pm 1$. As a result of the transformation introduced above, we have 
\begin{equation}\label{fv}
F(y)=\frac{(\rho_h-2)y(y-4)+7\rho_h-6}{(\rho_h-2)y(y^2-3y+3)+7\rho_h+2}(1+y), \quad 
  V_\epsilon(y) = \frac{(1-y)^2}{4}F(y)\left[\epsilon (1-y)\dot{F}(y)+\ell(\ell+1)\right].
\end{equation}
\begin{table}%[ht]
\small\setlength{\tabcolsep}{4.5pt}
\caption{Classification of the points $y=\pm 1$ for the relevant functions defined by   (\ref{S2onone})--(\ref{S0onone}), and (\ref{fv}),. The abbreviations $z$ ord $n$ and $p$ ord $m$ stand for zero of order $n$ and pole of order $m$, respectively.}
\begin{center}
\begin{tabular}{ | c | c | c | c | c | c | c | c }
\hline
$y$  & $F(y)$  & $V_\epsilon(y)$ & $S_2(y)$ & $S_{1}(y)$ & $S_{0}(y)$\\ \hline
$-1$ & z \mbox{ord} 1 & z \mbox{ord} 1 & z \mbox{ord} 4& z \mbox{ord} 3 & z \mbox{ord} 3 \\ \hline
$+1$ & $+1$  & z \mbox{ord} 2 & $4$ & p \mbox{ord} 2 & p \mbox{ord} 2\\ \hline
\end{tabular}
\label{tableEinsnone}
\end{center}
\end{table}
Table~\ref{tableEinsnone} indicates that the coefficients of the differential equation \eqref{ODEynone} share a common zero of order $2$ at $y = -1$ while $y = 1$ is a pole of order $4$ for the coefficient $S_0$. Hence, in order to apply the spectral method, we need to multiply \eqref{ODEynone} by $(1-y)^2/(1+y)^3$. As a result, we end up with the following differential equation
\begin{equation}\label{ODEhynone}
M_2(y)\ddot{\Phi}_{\Omega\ell\epsilon}(y) + M_1(y)\dot{\Phi}_{\Omega\ell\epsilon}(y) + M_0(y)\Phi_{\Omega\ell\epsilon}(y) = 0,
\end{equation}
where
\begin{equation}\label{S210honone}
M_2(y) = \frac{(1-y)^2}{1+y} F^2(y), \qquad
M_1(y) = i\Omega N_1(y)+N_0(y), \qquad
M_0(y) = \Omega^2 C_2(y)+i\Omega C_1(y)+C_0(y)
\end{equation}
with
\begin{eqnarray}
N_1(y) &=&\frac{2F^2(y)}{(1+y)^2}\left[4\rho_h+2(2-\rho_h)(1-y)-(a\rho_h+2)(1-y)^2\right],\label{N1}\\
N_0(y) &=&\frac{1-y}{1+y}\left[(1-y)\dot{F}(y)-2F(y)\right],\label{N0}\\
C_2(y) &=& \frac{4\rho_h^2}{(1+y)(1-y)^2}-\frac{F^2(y)}{(1+y)^3(1-y)^2}\left\{2(1-y^2)-\rho_h[ay^2-2y(1+a)+a-2]\right\}^2,\label{C2}\\
C_1(y) &=&\frac{F(y)}{(1+y)^3}\left\{(1+y)\dot{F}(y)\left[4\rho_h+2(2-\rho_h)(1-y)-(a\rho_h+2)(1-y)^2\right]\right.\nonumber\\
&&\left.-F(y)\left[8-(3+y)(1-y)(a\rho_h+2)\right]\right\},\label{C1}\\
C_0(y) &=& -\frac{4V_\epsilon(y)}{(1+y)(1-y)^2}.\label{C0}
\end{eqnarray}
It can be easily checked with Maple that
\begin{eqnarray}
    &&\lim_{y\to 1^{-}}M_2(y)=0=\lim_{y\to -1^{+}}M_2(y),\\
    &&\lim_{y\to 1^{-}}M_1(y)=2i\rho_h\Omega,\quad
    \lim_{y\to -1^{+}}M_1(y)=i\Omega\Lambda_1+\Lambda_0,\\
    &&\lim_{y\to 1^{-}}M_0(y)=A_2\Omega^2 +A_0,\quad
     \lim_{y\to -1^{+}}M_0(y)=B_2\Omega^2+i\Omega B_1+B_0,
\end{eqnarray}
where
\begin{eqnarray}
\Lambda_1 &=&4\rho_h-3\rho_h^2,\quad
\Lambda_0 = \frac{9\rho_h^2}{4}-6\rho_h+4,\quad
A_2=\frac{2\rho_h^2+12\rho_h-16}{3\rho_h-4},\label{coef1}\\
A_0&=&-\frac{\ell(\ell+1)}{2},\quad
B_2=\frac{\rho_h(9\rho_h^3-9\rho_h^2-18\rho_h+16)}{3\rho_h-4},\label{coef2}\\
B_1&=&\frac{9}{4}\rho_h^2(\rho_h-1)-\frac{9}{2}\rho_h+4,\quad
B_0=-\frac{9}{8}\left(\rho_h-\frac{4}{3}\right)\left[\left(\rho_h-\frac{4}{3}\right)\epsilon + \frac{2\ell(\ell+1)}{3}\right].\label{coef3}
\end{eqnarray}
In the final step leading to the application of the spectral method, we recast the differential equation \eqref{ODEhynone} into the following form
\begin{equation}\label{TSCH}
  L_0\left[\Phi_{\Omega\ell\epsilon}, \dot{\Phi}_{\Omega\ell\epsilon}, \ddot{\Phi}_{\Omega\ell\epsilon}\right] +  i L_1\left[\Phi_{\Omega\ell\epsilon}, \dot{\Phi}_{\Omega\ell\epsilon}, \ddot{\Phi}_{\Omega\ell\epsilon}\right]\Omega +  L_2\left[\Phi_{\Omega\ell\epsilon}, \dot{\Phi}_{\Omega\ell\epsilon}, \ddot{\Phi}_{\Omega\ell\epsilon}\right]\Omega^2 = 0
\end{equation}
with
\begin{eqnarray}
L_0\left[\Phi_{\Omega\ell\epsilon}, \dot{\Phi}_{\Omega\ell\epsilon}, \ddot{\Phi}_{\Omega\ell\epsilon}\right] &=& L_{00}(y)\Phi_{\Omega\ell\epsilon} + L_{01}(y)\dot{\Phi}_{\Omega\ell\epsilon} + L_{02}(y)\ddot{\Phi}_{\Omega\ell\epsilon},\label{L0none}\\
L_1\left[\Phi_{\Omega\ell\epsilon}, \dot{\Phi}_{\Omega\ell\epsilon}, \ddot{\Phi}_{\Omega\ell\epsilon}\right] &=&L_{10}(y)\Phi_{\Omega\ell\epsilon} + L_{11}(y)\dot{\Phi}_{\Omega\ell\epsilon} + L_{12}(y)\ddot{\Phi}_{\Omega\ell\epsilon}, \label{L1none}\\
L_2\left[\Phi_{\Omega\ell\epsilon}, \dot{\Phi}_{\Omega\ell\epsilon}, \ddot{\Phi}_{\Omega\ell\epsilon}\right] &=&L_{20}(y)\Phi_{\Omega\ell\epsilon} + L_{21}(y)\dot{\Phi}_{\Omega\ell\epsilon} + L_{22}(y)\ddot{\Phi}_{\Omega\ell\epsilon}.\label{L2none}
\end{eqnarray}
Moreover, in Table~\ref{tableZweinone}, we have summarized the $L_{ij}$ appearing in (\ref{L0none})--(\ref{L2none}) and their limiting values at $y = \pm 1$.

\begin{table}%[ht]
\small\setlength{\tabcolsep}{4.5pt}
\caption{Definitions of the coefficients $L_{ij}$ and their corresponding behaviours at the endpoints of the interval $-1 \leq y \leq 1$. The symbols appearing in this table have been defined in (\ref{coef1})-(\ref{coef3}).}
\begin{center}
\begin{tabular}{ | c | c | c | c | c | c | c | c }
\hline
$(i,j)$  & $\displaystyle{\lim_{y\to -1^+}}L_{ij}$  & $L_{ij}$ & $\displaystyle{\lim_{y\to 1^-}}L_{ij}$  \\ \hline
$(0,0)$ &  $B_0$          & $C_0$                  & $A_0$\\ \hline
$(0,1)$ &  $\Lambda_0$    & $N_0$                  & $0$\\ \hline
$(0,2)$ &  $0$            & $M_2$                  & $0$\\ \hline 
$(1,0)$ &  $B_1$          & $C_1$                  & $0$\\ \hline 
$(1,1)$ &  $\Lambda_1$    & $N_1$                  & $2\rho_h$\\ \hline 
$(1,2)$ &  $0$            & $0$                    & $0$\\ \hline 
$(2,0)$ &  $B_2$          & $C_2$                  & $A_2$\\ \hline
$(2,1)$ &  $0$            & $0$                    & $0$\\ \hline
$(2,2)$ &  $0$            & $0$                    & $0$\\ \hline
\end{tabular}
\label{tableZweinone}
\end{center}
\end{table} 

\subsection{The extreme case}
In this configuration, the metric function $g_{00}$ develops a double zero at $r = r_e$ when the parameter $\gamma$ reaches a critical value $32/27$. Introducing the rescaling $\rho=r/M$ and imposing the conditions $F(\rho_e) = 0$ and $F'(\rho_e) = 0$, we find that $\gamma = 32/27$ and $\rho_e = 4/3$. Introducing the dimensionless radial coordinate $x = \rho/\rho_e$ and substituting the aforementioned expressions into \eqref{ourODE}, we end up with the following spectral problem
\begin{equation}\label{ourODEext}
F(x)\frac{d}{dx}\left(F(x)\frac{d\psi_{\Omega\ell\epsilon}}{dx}\right)+\left[\frac{16}{9}\Omega^2-V_\epsilon(x)\right]\psi_{\Omega\ell\epsilon}(x)=0,\quad
F(x)=\frac{(2x+1)(x-1)^2}{2x^3+1},\quad
\Omega=M\omega
\end{equation} 
with effective potential given by \eqref{Veff}. To derive the QNM boundary conditions at the event horizon and at spatial infinity, we first analyse the asymptotic behaviour of the radial solution $\psi_{\Omega\ell\epsilon}$ in the limits $x \to 1^{+}$ and $x \to +\infty$. This asymptotic analysis enables us to determine the appropriate QNM boundary conditions corresponding to purely ingoing waves at the horizon and outgoing waves at infinity. Returning to equation \eqref{ourODEext} and expressing it in the form of \eqref{ODEZ}, a straightforward expansion around $x = 1$ reveals the leading-order behavior of the functions $p(x)$ and $q(x)$ near the horizon
\begin{equation}
p(x)=\frac{2}{x-1}+\mathcal{O}(1),\quad
\rev{q(x)}=\frac{16\Omega^2}{9(x - 1)^4} + \frac{128\Omega^2}{27(x - 1)^3}+\frac{A}{(x-1)^2}+\frac{B}{x-1}+\mathcal{O}(1).
\end{equation}
The expressions for the subleading coefficients $A$ and $B$ are omitted, as they are not relevant for the forthcoming analysis. Because $q(x)$ has a fourth-order pole at $x = 1$, it follows that $x = 1$ is an irregular singularity, rendering Frobenius's theory inapplicable in this scenario. On the other hand, since for $k = 1$ we have
\begin{equation}
    (x-1)^{k+1}p(x)=\mathcal{O}(x-1),\quad 
    (x-1)^{2k+2}q(x)=\frac{16\Omega^2}{9}+\mathcal{O}(x-1),
\end{equation}
then, according to \cite{Bender1999}, $x = 1$ is an irregular singular point of rank one. Consequently, the leading behaviour of the solutions to equation \eqref{ODEZ} in a neighbourhood of the event horizon can be deduced using the method outlined in \cite{Olver1994MAA}. To this purpose, we start by observing that by means of the transformation $\tau = (x-1)^{-1}$ mapping the event horizon at infinity and infinity to zero, \eqref{ODEZ} becomes
\begin{eqnarray}
&&\frac{d^2\psi_{\Omega\ell\epsilon}}{d\tau^2}+\mathfrak{C}(\tau)\frac{d\psi_{\Omega\ell\epsilon}}{d\tau}+\mathfrak{D}(\tau)\psi_{\Omega\ell\epsilon}(\tau)=0,\label{ODEZe}\\
&&\mathfrak{C}(\tau)=\mathcal{O}\left(\frac{1}{\tau^2}\right),\quad
\mathfrak{D}(\tau)=\frac{16\Omega^2}{9} + \frac{128\Omega^2}{27\tau}+\mathcal{O}\left(\frac{1}{\tau^2}\right).
\end{eqnarray}
A formal solution to \eqref{ODEZe} is given by \cite{Olver1994MAA}
\begin{equation}
\psi^{(\pm)}_{\Omega\ell\epsilon}(\tau)=\tau^{\mu_\pm}e^{\lambda_\pm \tau}\sum_{\kappa=0}^\infty\frac{\mathfrak{a}_{\kappa,\pm}}{\tau^\kappa},
\end{equation}
where $\lambda_\pm$, and $\mu_\pm$ can be computed according to \eqref{chareqns}. A straightforward computation shows that
\begin{equation}\label{lmu}
\lambda_\pm=\pm\frac{4}{3}i\Omega,\quad
\mu_\pm=\pm\frac{16}{9}i\Omega.
\end{equation}
At this point, it is important to observe that a radial field exhibiting purely ingoing behaviour near the event horizon ($x \to 1^+$) transforms, under the coordinate change $\tau = (x - 1)^{-1}$, into an outward-propagating mode as $\tau \to +\infty^{-}$. This correspondence supports the selection of the positive sign in the expressions for the characteristic exponents. Accordingly, the appropriate QNM boundary condition at the event horizon $x = 1$ takes the form
\begin{equation}\label{QNMBCe1}
\psi_{\Omega\ell\epsilon} \underset{{x \to 1^+}}{\longrightarrow} (x - 1)^{-\frac{16}{9}i\Omega} \exp\left(\frac{4i\Omega}{3(x - 1)}\right).
\end{equation}
By means of the transformation $\eta = 1/x$, it is not difficult to verify that the point at infinity is again an irregular singular point of rank one. Therefore, in the extremal case, the asymptotic behaviour of the solutions to equation \eqref{ourODEext} can be derived according to the method outlined in \cite{Olver1994MAA}. To this purpose, we  observe that
\begin{equation}
    p(x)=\mathcal{O}\left(\frac{1}{x^2}\right),\quad
    q(x)=\frac{16\Omega^2}{9}+\frac{16\Omega^2}{3x}+\mathcal{O}\left(\frac{1}{x^2}\right).
\end{equation}
With the help of \eqref{chareqns}, we immediately find that the QNM boundary condition at space-like infinity can be expressed as
 \begin{equation}\label{QNMBCzinfe}
    \psi_{\Omega\ell\epsilon}\underset{{x\to +\infty}}{\longrightarrow} x^{2i\Omega}e^{\frac{4}{3}i\Omega x}.
    \end{equation}
For the extremal problem we extract the formal asymptotic factors in the following way
\begin{eqnarray}
\psi_{\Omega\ell\epsilon}(x)&=&x^{\frac{34}{9}i\Omega}(x-1)^{-\frac{16}{9}i\Omega}e^{i\Omega\eta(x)} \Phi_{\Omega\ell\epsilon}(x),\quad
\eta(x)=\frac{4[(x-1)^2+1]}{3(x-1)}.
\end{eqnarray}
As in the non-extremal case, finite endpoint regularity of the remainder does not itself enforce a physical pole condition on the NIA. If we replace it into \eqref{ourODEext}, we end up with the differential equation
\begin{equation}\label{ODEzext}
P_{2e}(x)\Phi^{''}_{\Omega\ell\epsilon}(x)+P_{1e}(x)\Phi^{'}_{\Omega\ell\epsilon}(x)+P_{0e}(x)\Phi_{\Omega\ell\epsilon}(x)=0
\end{equation}
with
\begin{eqnarray}
P_{2e}(x)&=&x^2(x-1)^2 F^2(x),\\
P_{1e}(x)&=&x(x-1)F(x)\left\{x(x-1)F^{'}(x)+2i\Omega F(x)\left[x(x-1)\eta^{'}(x)+2x-\frac{34}{9}\right]\right\},\\
P_{0e}(x)&=&Q(x)\Omega^2+i\Omega L(x)-x^2(x-1)^2 V_\epsilon(x),\\
Q(x)&=&-F^2(x)\left[x(x-1)\eta^{'}(x)+2x-\frac{34}{9}\right]^2+\frac{16}{9}x^2(x-1)^2,\\
L(x)&=&x^2(x-1)^2 F(x)\left[F(x)\eta^{''}(x)+F^{'}(x)\eta^{'}(x)\right]+2x(x-1)\left(x-\frac{17}{9}\right)F(x)F^{'}(x)\nonumber\\
&&-2\left(x^2-\frac{34}{9}x+\frac{17}{9}\right)F^2(x).
\end{eqnarray}
As already done in the nonextremal case, we introduce the transformation $x = 2/(1-y)$. Furthermore, a dot denotes differentiation with respect to the new variable $y$. Then, equation \eqref{ODEzext} becomes
\begin{equation}\label{ODEye}
    S_{2e}(y)\ddot{\Phi}_{\Omega\ell\epsilon}(y)+S_{1e}(y)\dot{\Phi}_{\Omega\ell\epsilon}(y)+S_{0e}(y)\Phi_{\Omega\ell\epsilon}(y)=0,
\end{equation}
where
\begin{eqnarray}
S_{2e}(y)&=&(1+y)^2 F^2(y),\label{S2oe}\\
S_{1e}(y)&=&2i\Omega\frac{1+y}{1-y}F^2(y)\left[(1-y^2)\dot{\eta}(y)+\frac{2}{9}(17y+1)\right]+\frac{(1+y)^2}{1-y}F(y)\left[(1-y)\dot{F}(y)-2F(y)\right],\label{S1oe}\\
S_{0e}(y)&=&\Omega^2\Sigma_{2e}(y)+i\Omega\Sigma_{1e}(y)+\Sigma_{0e}(y)\label{S0oe}
\end{eqnarray}
with
\begin{eqnarray}
\Sigma_{2e}(y)&=&\frac{64(1+y)^2}{9(1-y)^4}-\frac{F^2(y)}{81(1-y)^2}\left[9(1-y^2)\dot{\eta}(y)+34y+2\right]^2,\\
\Sigma_{1e}(y)&=&\frac{(1+y)^2}{1-y}F(y)\left\{F(y)[(1-y)\ddot{\eta}(y)-2\dot{\eta}(y)]+(1-y)\dot{F}(y)\dot{\eta}(y)\right\}+\nonumber\\
&&\frac{34y^2+36y+2}{9(1-y)}F(y)\dot{F}(y)-\frac{34y^2+68y-30}{9(1-y)^2}F^2(y),\\
\Sigma_{0e}(y)&=&-\frac{4(1+y)^2}{(1-y)^4}V_\epsilon(y).
\end{eqnarray}
and the requirement that $\Phi_{\Omega\ell\epsilon}(y)$ is regular at $y = \pm 1$. As a result of the transformation introduced above, we have 
\begin{equation}\label{fve}
F(y)=\frac{(y-5)(1 + y)^2}{y^3-3y^2+3y-17}, \qquad 
\eta(y) = \frac{8(1+y^2)}{3(1-y^2)},
\end{equation}
while $V_\epsilon(y)$ is formally given by \eqref{fv}.
\begin{table}%[ht]
\small\setlength{\tabcolsep}{4.5pt}
\caption{Classification of the points $y = \pm 1$ for the relevant functions entering in (\ref{S2oe}), (\ref{S1oe}), and (\ref{S0oe}). The abbreviations $z$ ord $n$ and $p$ ord $m$ stand for zero of order $n$ and pole of order $m$, respectively.}
\begin{center}
\begin{tabular}{ | l | l | l | l |l |l | l | l}
\hline
$y$  & $F(y)$  & $V_\epsilon(y)$ & $\eta(y)$ & $S_{2e}(y)$ & $S_{1e}(y)$ & $S_{0e}(y)$\\ \hline
$-1$ & z \mbox{ord} 2 & z \mbox{ord} 2 & p \mbox{ord} 1 & z \mbox{ord} 6& z \mbox{ord} 4 & z \mbox{ord} 4 \\ \hline
$+1$ & $+1$  & z \mbox{ord} 2 & p \mbox{ord} 1 & $+4$ & p \mbox{ord} 2 & p \mbox{ord} 2\\ \hline
\end{tabular}
\label{table3}
\end{center}
\end{table}
Table~\ref{table3} indicates that the coefficients of the differential equation \eqref{ODEye} share a common zero of order $4$ at $y = -1$ while $y = 1$ is a pole of order $2$ for the coefficients $S_{1e}(y)$ and $S_{0e}(y)$. Hence, to apply the spectral method, we need to multiply \eqref{ODEye} by $(1-y)^2/(1+y)^4$. As a result, we end up with the following differential equation
\begin{equation}\label{ODEhynonee}
    M_{2e}(y)\ddot{\Phi}_{\Omega\ell\epsilon}(y)+M_{1e}(y)\dot{\Phi}_{\Omega\ell\epsilon}(y)+M_{0e}(y)\Phi_{\Omega\ell\epsilon}(y)=0,
\end{equation}
where
\begin{equation}\label{S210hononee}
  M_{2e}(y)=\left(\frac{1-y}{1+y}\right)^2 F^2(y),\quad
  M_{1e}(y)=i\Omega N_{1e}(y)+N_{0e}(y),\quad
  M_{0e}(y)=\Omega^2 C_{2e}(y)+i\Omega C_{1e}(y)+C_{0e}(y)
\end{equation}
with
\begin{eqnarray}
N_{1e}(y)&=&\frac{2(1-y)}{(1+y)^3}F^2(y)\left[(1-y^2)\dot{\eta}(y)+\frac{34}{9}y+\frac{2}{9}\right],\quad
N_{0e}(y)=\frac{1-y}{(1+y)^2}\left[(1-y)\dot{F}(y)-2F(y)\right],\label{N0e}\\
C_{2e}(y)&=&\frac{1}{(1+y)^4}\left\{\frac{64}{9}\left(\frac{1+y}{1-y}\right)^2-F^2(y)\left[(1-y^2)\dot{\eta}(y)+\frac{34}{9}y+\frac{2}{9}\right]^2\right\},\label{C2e}\\
C_{1e}(y)&=&\frac{1-y}{(1+y)^2}F(y)\left\{F(y)\left[(1-y)\ddot{\eta}(y)-2\dot{\eta}(y)\right]+(1-y)\dot{F}(y)\dot{\eta}(y)\right\}+\nonumber\\
&&\frac{(1-y)(34y^2+36y+2)}{9(1+y)^4}F(y)\dot{F}(y)-\frac{(34y^2+68y-30)F^2(y)}{9(1+y)^4},\label{C1e}\\
C_{0e}(y)&=&-\frac{4V_\epsilon(y)}{(1-y^2)^2}.\label{C0e}
\end{eqnarray}
It can be easily checked with Maple that
\begin{eqnarray}
&&\lim_{y\to 1^{-}}M_{2e}(y)=0=\lim_{y\to -1^{+}}M_{2e}(y),\quad 
\lim_{y\to 1^{-}}M_{1e}(y)=\frac{4}{3}i\Omega,\quad
\lim_{y\to -1^{+}}M_{1e}(y)=-\frac{4}{3}i\Omega,\\
&&\lim_{y\to 1^{-}}M_{0e}(y)=\frac{110}{27}\Omega^2-\frac{\ell(\ell+1)}{4},\quad
\lim_{y\to -1^{+}}M_{0e}(y)=\frac{356}{81}\Omega^2-\frac{\ell(\ell+1)}{4}.
\end{eqnarray}
Finally, in order to apply the spectral method, we rewrite the differential equation \eqref{ODEhynonee} into the following form
\begin{equation}\label{TSCHe}
\widehat{L}^{(e)}_0\left[\Phi_{\Omega\ell\epsilon},\dot{\Phi}_{\Omega\ell\epsilon},\ddot{\Phi}_{\Omega\ell\epsilon}\right]+ i\widehat{L}^{(e)}_1\left[\Phi_{\Omega\ell\epsilon},\dot{\Phi}_{\Omega\ell\epsilon},\ddot{\Phi}_{\Omega\ell\epsilon}\right]\Omega+ \widehat{L}_2^{(e)}\left[\Phi_{\Omega\ell\epsilon},\dot{\Phi}_{\Omega\ell\epsilon},\ddot{\Phi}_{\Omega\ell\epsilon}\right]\Omega^2=0
\end{equation}
with
\begin{eqnarray}
\widehat{L}^{(e)}_0\left[\Phi_{\Omega\ell\epsilon},\dot{\Phi}_{\Omega\ell\epsilon},\ddot{\Phi}_{\Omega\ell\epsilon}\right]&=&\widehat{L}^{(e)}_{00}(y)\Phi_{\Omega\ell\epsilon}+\widehat{L}^{(e)}_{01}(y)\dot{\Phi}_{\Omega\ell\epsilon}+\widehat{L}^{(e)}_{02}(y)\ddot{\Phi}_{\Omega\ell\epsilon},\label{L0nonee}\\
\widehat{L}^{(e)}_1\left[\Phi_{\Omega\ell\epsilon},\dot{\Phi}_{\Omega\ell\epsilon},\ddot{\Phi}_{\Omega\ell\epsilon}\right]&=&\widehat{L}^{(e)}_{10}(y)\Phi_{\Omega\ell\epsilon}+\widehat{L}^{(e)}_{11}(y)\dot{\Phi}_{\Omega\ell\epsilon}+\widehat{L}^{(e)}_{12}(y)\ddot{\Phi}_{\Omega\ell\epsilon},\label{L1nonee}\\
\widehat{L}^{(e)}_2\left[\Phi_{\Omega\ell\epsilon},\dot{\Phi}_{\Omega\ell\epsilon},\ddot{\Phi}_{\Omega\ell\epsilon}\right]&=&\widehat{L}^{(e)}_{20}(y)\Phi_{\Omega\ell\epsilon}+\widehat{L}^{(e)}_{21}(y)\dot{\Phi}_{\Omega\ell\epsilon}+\widehat{L}^{(e)}_{22}(y)\ddot{\Phi}_{\Omega\ell\epsilon}.\label{L2nonee}
\end{eqnarray}
Moreover, in Table~\ref{table4}, we have summarized the $\widehat{L}^{(e)}_{ij}$ appearing in (\ref{L0nonee})--(\ref{L2nonee}) and their limiting values at $y = \pm 1$.

\begin{table}%[ht]
\small\setlength{\tabcolsep}{4.5pt}
\caption{Definitions of the coefficients $\widehat{L}^{(e)}_{ij}$ and their corresponding behaviours at the endpoints of the interval $-1\leq y\leq 1$. }
\begin{center}
\begin{tabular}{ | l | l | l | l |l |l | l | l}
\hline
$(i,j)$  & $\displaystyle{\lim_{y\to -1^+}}\widehat{L}^{(e)}_{ij}$  & $\widehat{L}^{(e)}_{ij}$ & $\displaystyle{\lim_{y\to 1^-}}\widehat{L}^{(e)}_{ij}$  \\ \hline
$(0,0)$ &  $-\ell(\ell+1)/4$       & $C_{0e}$                  & $-\ell(\ell+1)/4$\\ \hline
$(0,1)$ &  $0$            & $N_{0e}$                  & $0$\\ \hline
$(0,2)$ &  $0$            & $M_{2e}$                  & $0$\\ \hline 
$(1,0)$ &  $0$            & $C_{1e}$                  & $0$\\ \hline 
$(1,1)$ &  $-4/3$ & $N_{1e}$                  & $4/3$\\ \hline 
$(1,2)$ &  $0$            & $0$                       & $0$\\ \hline 
$(2,0)$ &  $356/81$       & $C_{2e}$                  & $110/27$\\ \hline
$(2,1)$ &  $0$            & $0$                       & $0$\\ \hline
$(2,2)$ &  $0$            & $0$                       & $0$\\ \hline
\end{tabular}
\label{table4}
\end{center}
\end{table}

\section{Numerical method}\label{sec:numerical_method}

To discretise Eqs.~\eqref{TSCH} and \eqref{TSCHe}, we expand their factorised remainders in Chebyshev polynomials \cite{Boyd2000,Trefethen2000,Fox1968}
\begin{equation}\label{ChebExpansion}
\Phi_S(y)=\sum_{k=0}^{N_{\mathrm C}}a_kT_k(y),\qquad T_k(y)=\cos(k\arccos y),\quad S\in\{\mathrm{ne},\mathrm e\}.
\end{equation}
Here, $N_{\mathrm C}$ is the polynomial degree, $N$ in an oscillatory table is the reported overtone index, whereas in an NIA table it is only a local row index. The production collocation points are
\begin{equation}\label{ChebRoots}
y_j=-\cos\frac{(2j+1)\pi}{2(N_{\mathrm C}+1)},\qquad j=0,\ldots,N_{\mathrm C}.
\end{equation}
The alternative extrema grid is
\begin{equation}\label{ChebExtrema}
y_j=-\cos\frac{j\pi}{N_{\mathrm C}},\qquad j=0,\ldots,N_{\mathrm C}.
\end{equation}
All reported Hayward frequencies use the roots grid. The resulting quadratic pencil is \cite{Tisseur2001}
\begin{equation}\label{QuadraticEigenvalueProblem}
Q(\Omega)\mathbf a=(\mathbf M_0+i\Omega\mathbf M_1+\Omega^2\mathbf M_2)\mathbf a=0,
\qquad \mathbf a=(a_0,\ldots,a_{N_{\mathrm C}})^{\mathrm T}.
\end{equation}
The real matrices have order $N_{\mathrm C}+1$. The original calculations use \texttt{polyeig} in \textsc{Matlab}, with matrices assembled in \textsc{Maple} at 200-digit precision and imported through Advanpix \cite{mct2015}. The reported order comparisons are $N_{\mathrm C}=160,180,200$ for the 200-order tables and $250,280,300$ for the 300-order tables. These are the selection checks reported for the existing catalogue. The tables do not include mode-by-mode drifts at these orders, independent precision sweeps, or Hayward eigenfunction arrays. Consequently, the displayed digits and the working precision are not certified error estimates, and no new Hayward convergence measurements are asserted here. We display $\Im\Omega<0$ and $\Re\Omega\geq0$. This selection convention cannot establish absence of unstable modes.

\subsection{Boundary-sector selection and physical pole classification}\label{sec:pole_classification}
Let $dr_*/dr=1/F$. A QNM is an isolated pole of the Green function continued from its retarded domain, with horizon-ingoing solution $f_-$ and infinity-outgoing solution $f_+$. Away from exceptional normalisations its denominator is the constant tortoise-coordinate Wronskian. On the NIA, the two physical lateral continuations must be specified
\begin{equation}\label{eq:lateral_determinants}
D_\pm(\sigma)=\lim_{\varepsilon\downarrow0}
W_{r_*}\!\left[f_-(r,\pm\varepsilon-i\sigma),f_+(r,\pm\varepsilon-i\sigma)\right],\qquad \sigma>0.
\end{equation}
At a proposed pole, one must establish an isolated zero of the appropriately normalised determinant on the chosen continuation, and check that numerator cancellation or a singular Jost normalisation does not remove the Green-function pole. A finite-pencil root or a zero of the branch-cut strength is not this test. The long-range asymptotically flat problem makes branch-cut contamination relevant \cite{Leaver1986PRD,FortunaVega2023EPJC}, and no statement excluding every possible NIA pole follows merely from the presence of the cut. The local nonselection mechanism can also be seen directly for Hayward. At a simple horizon, $r_*\sim(2\kappa_h)^{-1}\log(r-r_h)$. At $\omega=-i\sigma$, division by the intended ingoing factor leaves the unwanted outgoing sector in the form
\begin{equation}\label{eq:hayward_unwanted_nonext}
\frac{e^{+i\omega r_*}}{e^{-i\omega r_*}}
\sim(r-r_h)^{\sigma/\kappa_h}.
\end{equation}
For any fixed integer $k$, an exponent $\sigma/\kappa_h>k$ permits a $C^k$ remainder, including the usual resonant logarithmic qualification. At infinity, the unwanted incoming sector divided by the outgoing one is $e^{-2\sigma r_*}$, which is flat under the algebraic compactification. In Schwarzschild, $\alpha=M\sigma$ gives $\sigma/\kappa_h=4\alpha$, the finite-regularity obstruction analysed in \cite{BaticDutykhSukaiti2026NIA}. A finite polynomial representation can approximate these unwanted sectors very accurately. Extraction of formal radiation factors followed by finite regularity therefore does not establish radiation-sector selection. At exact extremality $F\sim(r-r_e)^2/L^2$, so $r_*\sim-L^2/(r-r_e)$, up to logarithmic terms. The unwanted horizon remainder instead has the leading behaviour
\begin{equation}\label{eq:hayward_unwanted_ext}
e^{2\sigma r_*}\sim \exp\!\left[-\frac{2\sigma L^2}{r-r_e}\right]\times\hbox{an algebraic factor}.
\end{equation}
It is flat as $r\downarrow r_e$. This direct local observation reinforces the need for an extremal pole test, and it does not classify the global Hayward spectrum. In particular, an extremal test cannot be replaced by simply setting $\kappa_h=0$ in Eq.~\eqref{eq:hayward_unwanted_nonext}.

\subsection{Independent Schwarzschild evidence and its scope}
\label{sec:independent_evidence}
The subsequent study \cite{BaticDutykhSukaiti2026NIA} examines axial $(s,\ell)=(2,2)$ Schwarzschild perturbations with both the present linear radial compactification and a quadratic one. It finds a 68-point C1 ladder in $15.0786158201\leq-\Im\Omega\leq31.8428207066$ that survives roots-to-extrema changes but is not reproduced as a convergent finite-frequency ladder by C2. Independently of either pencil, horizon Jaff\'e and infinity Leaver--Tricomi-$U$ solutions yield two physical lateral Jost determinants that remain stably nonzero at all 68 C1 candidates. The same method recovers known off-axis overtones $n_{\rm QNM}=60,100,130$ at comparable damping, with regularised determinant magnitudes of order $10^{-48}$--$10^{-56}$. This is strong numerical evidence against the physical-pole interpretation of that tested ladder. The sampled scans and adjoining-contour winding tests in that work are not interval-certified exclusion theorems. The same study reports coefficient-tail and endpoint diagnostics and phase-aligned coefficient-vector comparisons. Its minimum roots/extrema overlap for the 68 candidates is $0.9999999999764556$, despite the nonzero lateral determinants. Such agreement verifies finite representations, not physical-pole character or continuum eigenfunction convergence. These results address the Schwarzschild benchmark only. No corresponding independent Jost calculation, alternative-compactification study, or eigenfunction convergence measurements for $\gamma>0$ or exact extremality are supplied here. Their NIA entries remain unclassified candidates. Conversely, the negative Schwarzschild result does not prove that all Hayward NIA candidates are spurious, nor that their weighted collective contribution converges to a branch cut. For future Hayward checks, a useful eigenfunction comparison is
\begin{equation}\label{eq:eigenfunction_diagnostic}
E_{N,N'}=\min_{\theta\in\mathbb R}\|\widehat\Phi_N-e^{i\theta}\widehat\Phi_{N'}\|_w,
\qquad \|\widehat\Phi_N\|_w=1,
\end{equation}
with a fixed stated physical coordinate, a common quadrature, and a specified norm. It should be accompanied by coefficient tails, residuals of the original radial equation at points not used in collocation, and analytic boundary-sector tests. Endpoint rows imposed in an extrema-grid pencil are not independent residual checks. We state these as outstanding diagnostics, not as computations already completed for the Hayward tables.

\section{Numerical results}
\label{sec:numerical_results}

The oscillatory Schwarzschild benchmarks and the low-lying scalar and electromagnetic comparisons \cite{Konoplya2022JCAP,Zainab2024EPL} test useful parts of the implementation. They do not validate the long NIA sequences. Throughout this section, \emph{candidate} denotes a finite-discretisation eigenvalue without an established physical-pole classification. Moreover, the NIA table index is not a physical overtone number. Frequencies absent from our selected lists are not thereby proved absent from the continuum spectrum.

\subsection{Scalar case}

For $\ell=0$, Table~\ref{scalarKonolpya01} agrees closely with \cite{Konoplya2022JCAP} through the first four Schwarzschild entries. At $\gamma=1$, agreement extends through $N=2$. The quoted near-axis and higher entries are not recovered by our selection procedure. Since the reference used independent checks, this remains an unresolved discrepancy, not evidence that those modes are spurious. Table~\ref{scalarKonolpya02} reproduces the listed $\ell=1$ comparison frequencies at $\gamma=0,1$ and supplies additional oscillatory candidates at intermediate parameters. The fundamental frequency generally becomes slightly more oscillatory and less damped as $\gamma$ increases. The NIA candidates in Tables~\ref{scalar01overdamped} and \ref{scalar02overdamped} have many gaps near $0.25$--$0.26$. Isolated low entries and enlarged gaps occur, notably at $\gamma=12/27$. These are features of the selected numerical lists, not established physical branch onsets. At $\gamma=1.18$, Tables~\ref{scalarg03Zainab} and \ref{scalarg03overdamped} give the near-extremal comparison. The $\ell=0$ semi-analytic reference differs by approximately $2\%$ in the real part and $10\%$ in damping, while the $\ell=1$ fundamental difference is below $1\%$. The short near-origin NIA pair in the latter sector requires separate validation.

\subsubsection{Extremal limit}
Tables~\ref{scalargextreme} and \ref{scalargextremeoverdamped} contain the extremal results. The monopole fundamental $\Omega=0.110697-0.087879i$ is close to the independently quoted WKB and time-domain values. Their componentwise differences are at most about $0.15\%$. At larger $\ell$, $\Re\Omega_0/(\ell+1/2)\simeq0.203$ and $|\Im\Omega_0|\simeq0.0815$. The NIA gaps and their finite-window variations are analysed in Sec.~\ref{sec:spacing_audit}. We do not assign them a physical overtone interpretation.

\subsection{Electromagnetic case}

Tables~\ref{emKonoplyaZeinab} and \ref{em01overdamped} show the non-extremal results. At $\gamma=0$, all seven listed oscillatory $\ell=1$ frequencies agree with \cite{Konoplya2022JCAP}. At $\gamma=1$ agreement extends through $N=4$. The two more strongly damped reference entries, including the near-axis entry, are not stable in the reported truncation sequence and require independent branch tracking. At $\gamma=1.1,1.18$, the fundamental frequencies agree at the sub-percent level with \cite{Zainab2024EPL}. Their damping decreases while their oscillation frequency increases modestly with deformation. Many NIA gaps are near $0.25$--$0.26$, but this is a statement about the finite candidate lists.

\subsubsection{Extremal limit}
Tables~\ref{emextreme} and \ref{emextremeoverdamped} give the extremal results. The fundamental oscillatory modes show the expected large-$\ell$ behaviour, with $\Re\Omega_0/(\ell+1/2)\simeq0.20$ and $|\Im\Omega_0|\simeq0.081$. At $\ell=1$, the listed NIA value $\Omega\simeq-4.23\times10^{-4}i$ is unclassified because it does not establish a slowly decaying physical mode. The two available gaps in that list are too irregular to define a regular spacing. At $\ell=2$, the displayed gap $1.514381$ is retained explicitly in the audit before applying the stated exclusion rule. Missing selected roots cannot be inferred to be missing physical modes.

\subsection{Axial effective-model results}\label{sec:axial_results}

Tables~\ref{tensor01} and \ref{tensor01overdamped} refer exclusively to $U_2^{\rm here}$ in Eq.~\eqref{eq:axial_here}. The Schwarzschild quadrupole $\Omega=0.373672-0.088962i$ is a useful common-limit benchmark. For $\ell=2$, increasing $\gamma$ to $1.18$ raises the fundamental real part by approximately $9\%$ and reduces its damping magnitude by approximately $22\%$ within this model. we underline that these percentages are not predictions independent of the effective-source closure. The listed NIA candidates have roughly quarter-spaced portions whose onset changes substantially with $\gamma$.

\begin{table}[htbp]
\small\setlength{\tabcolsep}{4.5pt}
\centering\small
\caption{Axial frequency comparison in $\Omega=M\omega$. Source-closure values are the eighth-order WKB values in Tables 1--4 of \cite{Bolokhov2026EPJC}; the present values use $U_2^{\rm here}$. Here, $\delta_R=100(\Re\Omega_{\rm here}/\Re\Omega_{\rm src}-1)$ and $\delta_I=100(|\Im\Omega_{\rm here}|/|\Im\Omega_{\rm src}|-1)$, in percent. These differences include both model dependence and residual method error.}
\label{tab:axial_closure_comparison}
\begin{tabular}{cccccrr}
\toprule
$\gamma$ & $\ell$ & $N$ & $\Omega_{\rm here}$ & $\Omega_{\rm src}$ & $\delta_R$ & $\delta_I$\\
\midrule
0 & 2 & 0 & $0.373672-0.088962i$ & $0.373669-0.088972i$ & $+0.001$ & $-0.011$\\
1.18 & 2 & 0 & $0.408845-0.069493i$ & $0.396195-0.074177i$ & $+3.19$ & $-6.31$\\
1.18 & 3 & 0 & $0.642615-0.076211i$ & $0.634564-0.078133i$ & $+1.27$ & $-2.46$\\
1.18 & 2 & 1 & $0.388809-0.212218i$ & $0.370157-0.225937i$ & $+5.04$ & $-6.07$\\
\bottomrule
\end{tabular}
\end{table}
Table~\ref{tab:axial_closure_comparison} compares matching $\gamma$, $\ell$, and overtone values. The reference sets $M=1$. The deformation-induced discrepancy is several percent for the quadrupole. The source-closure time-domain fundamental at $\gamma=1.18$, $\ell=2$ is $0.396224-0.074198i$ \cite{Bolokhov2026EPJC}, close to its WKB estimate. Thus, the large difference from the present value cannot be treated simply as disagreement between solvers for one potential. \cite{Malik2025IJTP} uses the same source-closure potential while the tabulated comparison frequencies are taken from \cite{Bolokhov2026EPJC}. We have not recomputed $U_2^{\rm src}$ with the present solver.

\subsubsection{Extremal limit}
Tables~\ref{tensorextreme} and \ref{tensorextremeoverdamped} give the extremal axial-model results. The fundamental oscillatory sequence tends towards $\Re\Omega_0/(\ell+1/2)\simeq0.20$ and $|\Im\Omega_0|\simeq0.08$. The NIA candidate gaps retain multipole dependence. For $\ell=3$, the gap $0.506707$ is almost twice its neighbours. Notice that this is a numerical list feature and not sufficient evidence for a missing overtone or a physical branch transition.

\subsection{Finite-table spacing statistics and uncertainty limits}

\label{sec:spacing_audit}
Let us reanalyse the printed frequencies from a different perspective. For $\alpha_j=-\Im\Omega_j$, we define $d_j=\alpha_{j+1}-\alpha_j$. Moreover, we only use pairs of displayed adjacent frequencies. A final printed gap whose next frequency is not displayed is not included. Within each list, we report the all-gap range, and call gaps with $d_j\leq1.5\,\operatorname{median}(d)$ \emph{retained} solely for a transparent descriptive comparison. Notice that this selection is not a pole or branch classifier. The extremal electromagnetic $\ell=1$ list is excluded from a regular-spacing summary because it has only two very irregular gaps. The last $K$ retained gaps define
\begin{equation}\label{eq:gap_statistics}
\bar d_K=\frac1K\sum_{j\in\mathcal W_K}d_j,\qquad
s_K=\left[\frac1{K-1}\sum_{j\in\mathcal W_K}(d_j-\bar d_K)^2\right]^{1/2}.
\end{equation}
Here, $s_K$ describes variation among gaps, not a standard error or numerical uncertainty. We recall that deterministic finite-window drift, discretisation error, conditioning, rounding, and physical-pole classification are distinct issues.

\begin{table}[htbp]
\small\setlength{\tabcolsep}{4.5pt}
\centering\small
\caption{Scalar-monopole gap statistics derived from displayed frequency pairs. $K_{\max}$ is the number of retained gaps, not modes. The final column is the sample gap dispersion over all retained gaps. It is not a solver error bar. The $\gamma=12/27$ gap at $N=0$ is excluded by the stated rule.}
\label{tab:scalar_gap_windows}
\begin{tabular}{ccrrrrr}
\toprule
$\gamma$ & $K_{\max}$ & $\bar d_3$ & $\bar d_5$ & $\bar d_7$ & $\bar d_{K_{\max}}$ & $s_{K_{\max}}$\\
\midrule
$0$ & 6 & 0.250027 & 0.250032 & --- & 0.250020 & $3.098\times10^{-5}$\\
$\frac{12}{27}$ & 5 & 0.253123 & 0.253110 & --- & 0.253110 & $7.649\times10^{-5}$\\
$1.18$ & 11 & 0.252743 & 0.252869 & 0.253022 & 0.253594 & $9.929\times10^{-4}$\\
$\frac{32}{27}$ & 7 & 0.252056 & 0.252058 & 0.252060 & 0.252060 & $4.121\times10^{-5}$\\
\bottomrule
\end{tabular}
\end{table}
At $\gamma=1.18$, the change from the seven-gap to three-gap mean is $2.79\times10^{-4}$, about $0.11\%$. For the extremal monopole, it is $3.24\times10^{-6}$. These quantify finite-window dependence, but they do not demonstrate asymptotic limit.

\begin{table}[htbp]
\small\setlength{\tabcolsep}{4.5pt}
\centering\small
\caption{Extremal candidate gap windows, using the last $K$ retained gaps. The product in the final column uses all $K_{\max}$ retained gaps. Dashes indicate insufficient retained gaps. The excluded gaps are $d_3=1.514381$ for electromagnetic $\ell=2$ and $d_2=0.506707$ for axial $\ell=3$. All other listed sequences retain all seven displayed adjacent pairs.}
\label{tab:extremal_gap_windows}
\begin{tabular}{ccrrrrr}
\toprule
Sector & $\ell$ & $K_{\max}$ & $\bar d_3$ & $\bar d_5$ & $\bar d_7$ & $(4/3)\bar d_{K_{\max}}$\\
\midrule
scalar & 0 & 7 & 0.252056 & 0.252058 & 0.252060 & 0.336079 \\
scalar & 1 & 7 & 0.252241 & 0.252247 & 0.252251 & 0.336335 \\
scalar & 2 & 7 & 0.252566 & 0.252586 & 0.252603 & 0.336804 \\
scalar & 3 & 7 & 0.253223 & 0.253274 & 0.253323 & 0.337764 \\
scalar & 4 & 7 & 0.254075 & 0.254165 & 0.254253 & 0.339004 \\
scalar & 5 & 7 & 0.256018 & 0.256191 & 0.256387 & 0.341849 \\
EM & 2 & 6 & 0.252400 & 0.252398 & --- & 0.336459 \\
EM & 3 & 7 & 0.252636 & 0.252643 & 0.252658 & 0.336877 \\
EM & 4 & 7 & 0.253143 & 0.253176 & 0.253212 & 0.337616 \\
EM & 5 & 7 & 0.254211 & 0.254271 & 0.254341 & 0.339122 \\
EM & 6 & 7 & 0.256679 & 0.256893 & 0.257129 & 0.342839 \\
axial model & 2 & 7 & 0.252285 & 0.252305 & 0.252303 & 0.336404 \\
axial model & 3 & 6 & 0.253117 & 0.253224 & --- & 0.337719 \\
axial model & 4 & 7 & 0.253746 & 0.253848 & 0.253949 & 0.338598 \\
axial model & 5 & 7 & 0.254435 & 0.254564 & 0.254706 & 0.339608 \\
axial model & 6 & 7 & 0.255358 & 0.255522 & 0.255720 & 0.340961 \\
axial model & 7 & 7 & 0.256641 & 0.256880 & 0.257149 & 0.342866 \\
\bottomrule
\end{tabular}
\end{table}

\begin{table}[htbp]
\small\setlength{\tabcolsep}{4.5pt}
\centering\small
\caption{Gap ranges before and after the descriptive selection, for lists with excluded gaps and representative unfiltered lists. The extremal EM $\ell=1$ row is not used in a regular-spacing mean.}
\label{tab:gap_exclusions}
\begin{tabular}{lcccl}
\toprule
Sector & $\gamma$ & $\ell$ & All-gap range & Retained-gap range\\
\midrule
scalar & $12/27$ & 0 & 0.252990--10.660423 & 0.252990--0.253190 \\
scalar & $32/27$ & 0 & 0.252013--0.252107 & 0.252013--0.252107 \\
scalar & $32/27$ & 5 & 0.255765--0.256883 & 0.255765--0.256883 \\
EM & $32/27$ & 1 & 1.766831--15.835615 & not defined \\
EM & $32/27$ & 2 & 0.251948--1.514381 & 0.251948--0.252844 \\
axial model & $32/27$ & 3 & 0.252981--0.506707 & 0.252981--0.253617 \\
axial model & $32/27$ & 7 & 0.256401--0.257943 & 0.256401--0.257943 \\
\bottomrule
\end{tabular}
\end{table}

For frequencies printed to $p$ decimal places and rounded to nearest, the display-rounding contribution to a difference is at most $10^{-p}$. For a mean of $K$ contiguous gaps, telescoping gives a rounding bound $10^{-p}/K$. After exclusions, a bound is $b\,10^{-p}/K$, where $b$ is the number of contiguous retained blocks. Thus, the bounds use $10^{-5}$ for the five-decimal scalar entries at $\gamma=0,12/27$, and $10^{-6}$ for the six-decimal entries, including the extremal tables. A common coarser precision gives a conservative bound when an endpoint pair has mixed display precision. Let us recall that these are only display-rounding bounds, not errors in the underlying eigenvalues. The observed departures from $1/4$ greatly exceed them in most deformed cases. The number-of-gaps sensitivity in these tables can be assessed from the published values. Stability against spectral order or arithmetic precision cannot be quantified from a single rounded list. A numerical error budget would require matched eigenvalues $\alpha_j^{(N_{\mathrm C},p)}$ and explicit order and precision variations. For an identified sequence, a bound $|\delta d_j|\leq |\delta\alpha_{j+1}|+|\delta\alpha_j|$ propagates those frequency errors. No such Hayward error bars are supplied here. The 200-digit working precision and the originally reported order checks do not replace that information. Moreover, even excellent finite-pencil error estimates would not establish a physical QNM spacing on the cut.

\subsection{Extremal length scale and limits of the spacing inference}
\label{sec:throat_limit}
The geometry itself gives the exact non-extremal surface gravity
\begin{equation}\label{HaywardSurfaceGravity}
M\kappa_h=\frac{3\rho_h-4}{4\rho_h},
\end{equation}
which tends from $1/4$ to zero as $\gamma$ tends from zero to $32/27$. At extremality, we have
\begin{equation}\label{HaywardExtremalExpansion}
F(\rho)=\frac9{16}(\rho-4/3)^2+\mathcal O((\rho-4/3)^3),
\end{equation}
and therefore
\begin{equation}\label{HaywardThroatLength}
F(r)=\frac{(r-r_e)^2}{L^2}+\mathcal O(((r-r_e)/L)^3),\qquad r_e=L=\frac{4M}{3}.
\end{equation}
Here, $L$ is the curvature radius of the near-horizon $\mathrm{AdS}_2$ factor, and not a finite proper length of the entire extremal throat. The local geometry is well defined independently of any interpretation of the NIA candidates. For this extremal Hayward metric, multiplying a dimensionless candidate gap by $L/M$ gives the exact conversion
\begin{equation}\label{HaywardExtremalDimensionlessProduct}
L\,\Delta(-\Im\omega)=\frac43\,d.
\end{equation}
Consequently, $d\simeq1/4$ and $L\Delta(-\Im\omega)\simeq1/3$ are the same empirical statement in different units. Table~\ref{tab:extremal_gap_windows} broadens the original monopole example to the available multipoles and sectors. Its full retained means give products from about $0.33608$ to $0.34287$, approximately $0.82\%$--$2.86\%$ above $1/3$. However, the electromagnetic $\ell=1$ data do not support a regular-spacing entry. The finite-window and multipole dependence, unknown discretisation errors, and unclassified pole character preclude a universal physical scaling claim. We therefore do not interpretat  that a throat scale has been shown to control Hayward QNM damping. Varying $M$ at fixed extremal dimensionless geometry would merely rescale frequencies and cannot provide an independent test of the coefficient $1/3$.

\section{Comparison across the three effective geometries}\label{sec:trilogy_comparison}

The three studies use a common effective axial model and related spectral discretisations \cite{Batic2026EPJC,Batic2026PRD}. Their oscillatory frequencies can be compared after conversion to $\widehat\omega=M\omega$, with the proviso that their deformation parameters are not physically matched. The fractional fundamental shifts are
\begin{equation}\label{TrilogyFractionalShifts}
\delta_R=\frac{\Re\widehat\omega_0-\Re\widehat\omega_0^{\rm Sch}}{\Re\widehat\omega_0^{\rm Sch}},\qquad
\delta_I=\frac{|\Im\widehat\omega_0|-|\Im\widehat\omega_0^{\rm Sch}|}{|\Im\widehat\omega_0^{\rm Sch}|}.
\end{equation}
The representative Planck-star configuration $(M,\gamma)=(1,9/2)$ has negative real-frequency shifts and increased damping for the standard scalar, electromagnetic, and axial fundamentals. The nearly extremal Bonanno--Reuter configuration $M=3.503$ and extremal Hayward configuration instead have unchanged or increased real frequencies and reduced damping. For example,
\begin{equation}\label{TrilogyQuadrupoleComparison}
\widehat\omega^{\rm BR}_{0,\ell=2}=0.418538-0.069121i,\qquad
\widehat\omega^{\rm H}_{0,\ell=2}=0.409045-0.069319i,
\end{equation}
compared with $\widehat\omega^{\rm Sch}_{0,\ell=2}=0.373672-0.088962i$. This numerical proximity does not imply isospectrality or an invariant map between the geometries. The NIA comparisons require a more substantial qualification. Similarity of ladders generated by related factorisations and polynomial trial spaces is not an independent continuum-pole test. The independent Schwarzschild result \cite{BaticDutykhSukaiti2026NIA} shows precisely why a regular quarter-spaced numerical sequence can survive stringent finite-pencil checks without furnishing a physical QNM family. We therefore astein ourselves from introducing a classification into surface-gravity-controlled and throat-controlled physical branches. Imaginary-axis sequences in \cite{Batic2026EPJC,Batic2026PRD} would also require geometry-specific analytic-continuation and pole tests before that interpretation could be defended. Nor does the present result prove their candidates are all branch-cut artifacts. Establishing a collective cut-response interpretation requires appropriately weighted resolvent or time-domain convergence, which is not available for these comparative data. For the same reason, using a value predicted from $1/(3L)$ as the extremal comparison point would not be independent numerical evidence for that relation. The Hayward descriptive data in Sec.~\ref{sec:spacing_audit} are therefore kept separate from geometric formulae and from the supported low-lying oscillatory comparisons.

\section{Conclusions and outlook}\label{sec:conclusions}

We have formulated separate non-extremal and exactly extremal Hayward radial problems and documented their multiprecision Chebyshev spectra. Most available low-lying scalar and electromagnetic oscillatory benchmarks are reproduced, while several strongly damped or near-axis literature frequencies remain unresolved by the reported selection procedure. The oscillatory results generally indicate modest increases in frequency and reduced damping with deformation for the representative fundamentals. The axial results refer specifically to $U_2^{\rm here}$, not to a unique gravitational perturbation theory of the Hayward background. The explicit potential difference and published-frequency comparison quantify this limitation. The imaginary-axis catalogue has a different evidentiary status. The subsequent axial Schwarzschild Jost calculation \cite{BaticDutykhSukaiti2026NIA} demonstrates that accurate, stable finite-pencil eigenpairs can fail a physical pole test. We accordingly retain the Hayward NIA entries as unclassified spectral candidates and do not claim that they establish additional QNM families, universal high-damping spacings, or throat-controlled relaxation. The local endpoint analysis explains why formal radiation factorisation and finite regularity cannot settle the question, including at the extremal double horizon. However, it does not rule out every possible purely imaginary Hayward QNM. The descriptive statistics state the gap-selection rule, finite-window dependence, and rounding limitations. They quantify what follows from the printed lists, but supply neither order/precision error bars nor eigenfunction convergence data. The extremal products near $1/3$ are a rescaling of the same roughly quarter-spaced candidates. We reiterate that their deviations and sector dependence do not support equality to numerical accuracy or a universal physical law.

Further validation requires reconstructing the Hayward eigenfunctions at different spectral resolutions and arithmetic precisions, evaluating residuals away from the collocation points, and testing alternative radial representations. Physical lateral Jost determinants, or an equivalent analytically controlled resonance criterion, should then be evaluated using off-axis modes at comparable damping as controls. The extremal geometry requires a separate analytic continuation. Interpreting the finite spectral nodes as a collective representation of the branch-cut response would additionally require demonstrating convergence of the weighted resolvent or the corresponding time-domain response. These checks remain outstanding for the present Hayward catalogue.

Even with this validation, pole locations alone would not determine excitation amplitudes or observational visibility. The selected lower-half-plane frequencies neither establish spectral completeness nor prove stability. They also provide no assessment of stability against polar perturbations, coupled metric–source perturbations, or nonlinear effects, and do not resolve the stability of the Cauchy horizon.

\subsection*{Code availability}

\noindent The codes used to assemble the spectral matrices and to post-process the quadratic eigenvalue problems are available at the following URL address
\begin{center}
  \url{https://github.com/dutykh/ASafeGravity/}
\end{center}

\subsection*{Funding}

This research received no specific grant from any funding agency in the public, commercial, or not-for-profit sectors.

\begin{table}%[ht]
\small\setlength{\tabcolsep}{4.5pt}
\centering
\caption{Oscillatory QNM benchmarks and additional spectral candidates for scalar perturbations of the non-extremal Hayward black hole for $\ell=0$, and different values of the parameter $\gamma$. The locations of the event horizon $\rho_h$ are given in Table~\ref{table:event}. The results are obtained using our spectral method with reported spectral truncation order $N_{\mathrm C}=300$ and 200-digit precision. Here, $\Omega$ denotes the dimensionless QNM frequency, and $N$ the overtone number. 'N/A' indicates data not available, and 'SM' refers to the Spectral Method.}
\label{scalarKonolpya01}
\vspace*{1em}
\begin{tabular}{||c|c|c|c|c|c|c|c|c|c|c|c|c||}
\hline\hline
$\gamma$   &$\ell$ & $N$ & $\Omega$ \cite{Konoplya2022JCAP} & $\Omega$ (SM) & $\gamma$  & $\ell$ & $N$ & $\Omega$ \cite{Konoplya2022JCAP} & $\Omega$ (SM) \\ [0.5ex]
\hline\hline
$0$    &$0$    & $0$ &$0.110455-0.104896i$            & $0.110455-0.104896i$          & $1$     & $0$    & $0$ & $0.113494-0.089160i$   & $0.113494-0.089160i$\\
       &       & $1$ &$0.086117-0.348053i$            & $0.086117-0.348052i$          &         &        & $1$ & $0.066731-0.319873i$   & $0.066731-0.319873i$\\
       &       & $2$ &$0.075742-0.601079i$            & $0.075742-0.601079i$          &         &        & $2$ & $0.041068-0.576924i$   & $0.041113-0.576958i$\\
       &       & $3$ &$0.070410-0.853678i$            & $0.070404-0.853672i$          &         &        & $3$ & $0.021679-0.833067i$   & N/A\\
       &       & $4$ &$0.067074-1.105630i$            & N/A                           &         &        & $4$ & $0.000000-1.082236i$   & N/A\\
       &       & $5$ &$0.064742-1.357140i$            & N/A                           &         &        & $5$ & $0.001449-1.317232i$   & N/A\\
$4/27$&$0$     & $0$ &$0.111600-0.103409i$            & $0.111600-0.103409i$          & $20/27$ & $0$    & $0$ & $0.114963-0.094678i$  & $0.114963-0.094678i$\\
       &       & $1$ &N/A                             & $0.087198-0.341961i$          &         &        & $1$ & N/A                   & $0.077295-0.307784i$\\
       &       & $2$ &N/A                             & $0.075182-0.590569i$          &         &        & $2$ & N/A                   & N/A\\
       &       & $3$ &N/A                             & $0.067323-0.839220i$          &         &        & $3$ & N/A                   & N/A\\
$8/27$&$0$     & $0$ &$0.112698-0.101699i$            & $0.112698-0.101700i$          & $24/27$ & $0$    & $0$ & $0.114675-0.091539i$  & $0.114675-0.091539i$\\
       &       & $1$ &N/A                             & $0.087656-0.335037i$          &         &        & $1$ & N/A                   & $0.055643-0.303334i$\\
       &       & $2$ &N/A                             & $0.072501-0.578756i$          &         &        & $2$ & N/A                   & $0.051504-0.566716i$\\
       &       & $3$ &N/A                             & $0.059532-0.823395i$          &         &        & $3$ & N/A                   & N/A\\
$12/27$&$0$    & $0$ &$0.113698-0.099717i$            & $0.113698-0.099717i$          & $28/27$ & $0$    & $0$ & $0.112827-0.088526i$    & $0.112827-0.088526i$\\
       &       & $1$ &N/A                             & $0.087031-0.327092i$          &         &        & $1$ & N/A                     & $0.068047-0.313819i$\\
       &       & $2$ &N/A                             & $0.065789-0.565546i$          &         &        & $2$ & N/A                     & $0.044449-0.563170i$\\
       &       & $3$ &N/A                             & N/A          &         &        & $3$ & N/A                              & N/A\\
[1ex]
\hline\hline 
\end{tabular}
\end{table}

\begin{table}%[ht]
\small\setlength{\tabcolsep}{4.5pt}
\centering
\caption{Unclassified purely imaginary spectral candidates for scalar perturbations of the non-extremal Hayward black hole for $\ell=0$, and different values of the parameter $\gamma$. The locations of the event horizon $\rho_h$ are given in Table~\ref{table:event}. The corresponding results are obtained through our spectral method, with reported spectral truncation order $N_{\mathrm C}=200$ with a precision of $200$ digits. In this context, $\Omega$ and $N$ represent the dimensionless frequency and the local row index, respectively, while $\Delta\Omega=\Omega_N-\Omega_{N+1}$. The notation 'SM' stands for Spectral Method. These entries have not been verified as isolated Green-function poles. The displayed precision is not a numerical error bound (see Secs.~\ref{sec:pole_classification} and \ref{sec:spacing_audit}).}
\label{scalar01overdamped}
\vspace*{1em}
\begin{tabular}{||c|c|c|c|c|c|c|c|c|c|c|c|c||}
\hline\hline
$\gamma$&$\ell$ & $N$ &$\Omega$ (SM) & $\Delta\Omega$ & $\gamma$     & $\ell$ & $N$ & $\Omega$ (SM) & $\Delta\Omega$ \\ [0.5ex]
\hline\hline
$0$    &$0$    & $0$ &$0.0000-27.86817i$   & $0.249958i$            & $1$     & $0$    & $0$ & $0.0000-4.628382i$    & $0.252667i$\\
       &       & $1$ &$0.0000-28.11813i$   & $0.250039i$            &         &        & $1$ & $0.0000-4.881049i$    & $0.252632i$\\
       &       & $2$ &$0.0000-28.36817i$   & $0.250042i$            &         &        & $2$ & $0.0000-5.133681i$    & $0.252606i$\\
       &       & $3$ &$0.0000-28.61821i$   & $0.250016i$            &         &        & $3$ & $0.0000-5.386287i$    & $0.252584i$\\
       &       & $4$ &$0.0000-28.86823i$   & $0.250037i$            &         &        & $4$ & $0.0000-5.638872i$    & $0.252566i$\\
       &       & $5$ &$0.0000-29.11827i$   & $0.250026i$            &         &        & $5$ & $0.0000-5.891437i$    & $0.252550i$\\
       &       & $6$ &$0.0000-29.36829i$   & $0.250028i$            &         &        & $6$ & $0.0000-6.143987i$    & $0.252536i$\\
$4/27$ &$0$    & $0$ &$0.0000-22.06990i$   & $0.252630i$            & $20/27$ & $0$    & $0$ & $0.0000-9.441956i$    & $0.252846i$\\
       &       & $1$ &$0.0000-22.32253i$   & $0.252651i$            &         &        & $1$ & $0.0000-9.694801i$    & $0.252859i$\\
       &       & $2$ &$0.0000-22.57518i$   & $0.252655i$            &         &        & $2$ & $0.0000-9.947660i$    & $0.252867i$\\
       &       & $3$ &$0.0000-22.82783i$   & $0.252650i$            &         &        & $3$ & $0.0000-10.20053i$    & $0.252864i$\\
       &       & $4$ &$0.0000-23.08048i$   & $0.252651i$            &         &        & $4$ & $0.0000-10.45339i$    & $0.252863i$\\
       &       & $5$ &$0.0000-23.33313i$   & $0.252651i$            &         &        & $5$ & $0.0000-10.70625i$    & $0.252861i$\\
$8/27$ &$0$    & $0$ &$0.0000-18.05494i$   & $0.253052i$            & $24/27$ & $0$    & $0$ & $0.0000-4.096256i$    & $2.557819i$\\
       &       & $1$ &$0.0000-18.30799i$   & $0.253052i$            &         &        & $1$ & $0.0000-6.654075i$    & $0.252680i$\\
       &       & $2$ &$0.0000-18.56104i$   & $0.253052i$            &         &        & $2$ & $0.0000-6.906755i$    & $0.252700i$\\
       &       & $3$ &$0.0000-18.81409i$   & $0.253051i$            &         &        & $3$ & $0.0000-7.159455i$    & $0.252689i$\\
       &       & $4$ &$0.0000-19.06714i$   & $0.253052i$            &         &        & $4$ & $0.0000-7.412144i$    & $0.252682i$\\
       &       & $5$ &$0.0000-19.32020i$   & $0.253051i$            &         &        & $5$ & $0.0000-7.664825i$    & $0.252675i$\\
       &       & $6$ &$0.0000-19.57325i$   & $0.253051i$            &         &        & $6$ & $0.0000-7.917501i$    & $0.252670i$\\
$12/27$&$0$    & $0$ &$0.0000-4.107617i$   & $10.66042i$            & $28/27$ & $0$    & $0$ & $0.0000-3.616384i$    & $0.252792i$\\
       &       & $1$ &$0.0000-14.76804i$   & $0.253187i$            &         &        & $1$ & $0.0000-3.869176i$    & $0.252717i$\\
       &       & $2$ &$0.0000-15.02123i$   & $0.252993i$            &         &        & $2$ & $0.0000-4.121893i$    & $0.252670i$\\
       &       & $3$ &$0.0000-15.27422i$   & $0.253154i$            &         &        & $3$ & $0.0000-4.374563i$    & $0.252628i$\\
       &       & $4$ &$0.0000-15.52738i$   & $0.253106i$            &         &        & $4$ & $0.0000-4.627191i$    & $0.252592i$\\
       &       & $5$ &$0.0000-15.78048i$   & $0.253111i$            &         &        & $5$ & $0.0000-4.879783i$    & $0.252563i$\\
       &       & $6$ &$0.0000-16.03359i$   & $0.253113i$            &         &        & $6$ & $0.0000-5.132346i$    & $0.252537i$\\
[1ex]
\hline\hline 
\end{tabular}
\end{table}

\begin{table}%[ht]
\small\setlength{\tabcolsep}{4.5pt}
\centering
\caption{Oscillatory QNM benchmarks and additional spectral candidates for scalar perturbations of the non-extremal Hayward black hole for $\ell=1$, and different values of the parameter $\gamma$. The locations of the event horizon $\rho_h$ are given in Table~\ref{table:event}. The results are obtained using our spectral method with reported spectral truncation order $N_{\mathrm C}=300$ and 200-digit precision. Here, $\Omega$ denotes the dimensionless QNM frequency, and $N$ the overtone number. 'N/A' indicates data not available, and 'SM' refers to the Spectral Method.}
\label{scalarKonolpya02}
\vspace*{1em}
\begin{tabular}{||c|c|c|c|c|c|c|c|c|c|c|c|c||}
\hline\hline
$\gamma$   &$\ell$ & $N$ & $\Omega$ \cite{Konoplya2022JCAP} & $\Omega$ (SM) & $\gamma$  & $\ell$ & $N$ & $\Omega$ \cite{Konoplya2022JCAP} & $\Omega$ (SM) \\ [0.5ex]
\hline\hline
$0$    &$1$    & $0$ &$0.292936-0.097660i$            & $0.292936-0.097660i$          & $1$     & $1$    & $0$ & $0.305627-0.085590i$  & $0.305627-0.085590i$\\
       &       & $1$ &$0.264449-0.306258i$            & $0.264449-0.306257i$          &         &        & $1$ & $0.274844-0.263717i$  & $0.274844-0.263717i$\\
       &       & $2$ &$0.229540-0.540134i$            & $0.229539-0.540133i$          &         &        & $2$ & $0.218494-0.464862i$  & $0.218494-0.464862i$\\
       &       & $3$ &$0.203259-0.788298i$            & $0.203258-0.788298i$          &         &        & $3$ & $0.158106-0.700610i$  & $0.158106-0.700610i$\\
       &       & $4$ &$0.185109-1.040762i$            & $0.185109-1.040762i$          &         &        & $4$ & $0.112793-0.957178i$  & $0.112793-0.957179i$\\
       &       & $5$ &$0.172077-1.294120i$            & $0.172077-1.294120i$          &         &        & $5$ & $0.081256-1.220106i$  & $0.081228-1.220130i$\\
$4/27$&$1$     & $0$ &N/A                             & $0.294756-0.096422i$          & $20/27$ & $1$    & $0$ & N/A                   & $0.302422-0.089765i$\\
       &       & $1$ &N/A                             & $0.267288-0.301696i$          &         &        & $1$ & N/A                   & $0.275909-0.277320i$\\
       &       & $2$ &N/A                             & $0.232851-0.530815i$          &         &        & $2$ & N/A                   & $0.233336-0.481810i$\\
       &       & $3$ &N/A                             & $0.205777-0.773847i$          &         &        & $3$ & N/A                   & $0.178810-0.703207i$\\
       &       & $4$ &N/A                             & $0.185919-1.021330i$          &         &        & $4$ & N/A                   & $0.130503-0.967073i$\\
       &       & $5$ &N/A                             & $0.170492-1.269909i$          &         &        & $5$ & N/A                   & $0.106309-1.208819i$\\
       &       & $6$ &N/A                             & $0.157635-1.518642i$          &         &        & $6$ & N/A                   & N/A\\
       &       & $7$ &N/A                             & $0.146265-1.767291i$          &         &        & $7$ & N/A                   & N/A\\
       &       & $8$ &N/A                             & $0.135736-2.015837i$          &         &        & $8$ & N/A                   & N/A\\
       &       & $9$ &N/A                             & $0.125605-2.264260i$          &         &        & $9$ & N/A                   & N/A\\
$8/27$&$1$     & $0$ &N/A                             & $0.296625-0.095044i$          & $24/27$ & $1$    & $0$ & N/A                   & $0.304307-0.087494i$\\
       &       & $1$ &N/A                             & $0.270009-0.296615i$          &         &        & $1$ & N/A                   & $0.275958-0.269511i$\\
       &       & $2$ &N/A                             & $0.235512-0.520418i$          &         &        & $2$ & N/A                   & $0.224865-0.469863i$\\
       &       & $3$ &N/A                             & $0.206571-0.757663i$          &         &        & $3$ & N/A                   & $0.166944-0.704460i$\\
       &       & $4$ &N/A                             & $0.183274-0.999506i$          &         &        & $4$ & N/A                   & $0.114317-0.956832i$\\
       &       & $5$ &N/A                             & $0.162879-1.242691i$          &         &        & $5$ & N/A                   & $0.092302-1.208775i$\\
       &       & $6$ &N/A                             & $0.143256-1.486339i$          &         &        & $6$ & N/A                   & N/A\\
       &       & $7$ &N/A                             & $0.122683-1.730364i$          &         &        & $7$ & N/A                   & N/A\\
$12/27$&$1$    & $0$ &N/A                             & $0.298538-0.093498i$          & $28/27$ & $1$    & $0$ & N/A                   & $0.306038-0.084920i$\\
       &       & $1$ &N/A                             & $0.272501-0.290919i$          &         &        & $1$ & N/A                   & $0.274286-0.261906i$\\
       &       & $2$ &N/A                             & $0.237101-0.508765i$          &         &        & $2$ & N/A                   & $0.216596-0.463142i$\\
       &       & $3$ &N/A                             & $0.204469-0.739532i$          &         &        & $3$ & N/A                   & $0.156316-0.698830i$\\
       &       & $4$ &N/A                             & $0.174337-0.975221i$          &         &        & $4$ & N/A                   & $0.110435-0.953070i$\\
       &       & $5$ &N/A                             & $0.142506-1.213121i$          &         &        & $5$ & N/A                   & N/A\\
       &       & $6$ &N/A                             & $0.101169-1.454444i$          &         &        & $6$ & N/A                   & N/A\\
[1ex]
\hline\hline 
\end{tabular}
\end{table}

\begin{table}%[ht]
\small\setlength{\tabcolsep}{4.5pt}
\centering
\caption{Unclassified purely imaginary spectral candidates for scalar perturbations of the non-extremal Hayward black hole for $\ell=1$, and different values of the parameter $\gamma$. The locations of the event horizon $\rho_h$ are given in Table~\ref{table:event}. The corresponding results are obtained through our spectral method, with reported spectral truncation order $N_{\mathrm C}=300$ with a precision of $200$ digits. In this context, $\Omega$ and $N$ represent the dimensionless frequency and the local row index, respectively, while $\Delta\Omega=\Omega_N-\Omega_{N+1}$. The notation 'SM' stands for Spectral Method. These entries have not been verified as isolated Green-function poles. The displayed precision is not a numerical error bound (see Secs.~\ref{sec:pole_classification} and \ref{sec:spacing_audit}).}
\label{scalar02overdamped}
\vspace*{1em}
\begin{tabular}{||c|c|c|c|c|c|c|c|c|c|c|c|c||}
\hline\hline
$\gamma$&$\ell$ & $N$ &$\Omega$ (SM) & $\Delta\Omega$ & $\gamma$     & $\ell$ & $N$ & $\Omega$ (SM) & $\Delta\Omega$ \\ [0.5ex]
\hline\hline
$0$    &$1$    & $0$ &$0.0000-28.07792i$  & $0.250172i$            & $1$     & $1$    & $0$ & $0.0000-4.577749i$    & $0.255512i$\\
       &       & $1$ &$0.0000-28.32809i$  & $0.250250i$            &         &        & $1$ & $0.0000-4.833261i$    & $0.255186i$\\
       &       & $2$ &$0.0000-28.57834i$  & $0.250159i$            &         &        & $2$ & $0.0000-5.088447i$    & $0.254910i$\\
       &       & $3$ &$0.0000-28.82850i$  & $0.250209i$            &         &        & $3$ & $0.0000-5.343357i$    & $0.254671i$\\
       &       & $4$ &$0.0000-29.07871i$  & $0.250187i$            &         &        & $4$ & $0.0000-5.598028i$    & $0.254463i$\\
       &       & $5$ &$0.0000-29.32890i$  & $0.250186i$            &         &        & $5$ & $0.0000-5.852491i$    & $0.254281i$\\
       &       & $6$ &$0.0000-29.57908i$  & $0.250189i$            &         &        & $6$ & $0.0000-6.106772i$    & $0.254122i$\\
       &       & $7$ &$0.0000-29.82927i$  & $0.250182i$            &         &        & $7$ & $0.0000-6.360895i$    & $0.253982i$\\
$4/27$ &$1$    & $0$ &$0.0000-22.05120i$  & $0.252810i$            & $20/27$ & $1$    & $0$ & $0.0000-9.415822i$    & $0.253553i$\\
       &       & $1$ &$0.0000-22.30401i$  & $0.252881i$            &         &        & $1$ & $0.0000-9.669375i$    & $0.253545i$\\
       &       & $2$ &$0.0000-22.55689i$  & $0.252862i$            &         &        & $2$ & $0.0000-9.922920i$    & $0.253517i$\\
       &       & $3$ &$0.0000-22.80975i$  & $0.252858i$            &         &        & $3$ & $0.0000-10.17644i$    & $0.253480i$\\
       &       & $4$ &$0.0000-23.06261i$  & $0.252853i$            &         &        & $4$ & $0.0000-10.42992i$    & $0.253448i$\\
       &       & $5$ &$0.0000-23.31546i$  & $0.252849i$            &         &        & $5$ & $0.0000-10.68337i$    & $0.253417i$\\
       &       & $6$ &$0.0000-23.56831i$  & $0.252843i$            &         &        & $6$ & $0.0000-10.93678i$    & $0.253389i$\\
       &       & $7$ &$0.0000-23.82116i$  & $0.252839i$            &         &        & $7$ & $0.0000-11.19017i$    & $0.253363i$\\
$8/27$ &$1$    & $0$ &$0.0000-18.29004i$  & $0.253318i$            & $24/27$ & $1$    & $0$ & $0.0000-6.618411i$    & $0.254110i$\\
       &       & $1$ &$0.0000-18.54336i$  & $0.253301i$            &         &        & $1$ & $0.0000-6.872522i$    & $0.254007i$\\
       &       & $2$ &$0.0000-18.79666i$  & $0.253290i$            &         &        & $2$ & $0.0000-7.126529i$    & $0.253894i$\\
       &       & $3$ &$0.0000-19.04995i$  & $0.253287i$            &         &        & $3$ & $0.0000-7.380422i$    & $0.253802i$\\
       &       & $4$ &$0.0000-19.30324i$  & $0.253278i$            &         &        & $4$ & $0.0000-7.634224i$    & $0.253720i$\\
       &       & $5$ &$0.0000-19.55652i$  & $0.253273i$            &         &        & $5$ & $0.0000-7.887944i$    & $0.253646i$\\
       &       & $6$ &$0.0000-19.80979i$  & $0.253267i$            &         &        & $6$ & $0.0000-8.141590i$    & $0.253578i$\\
       &       & $7$ &$0.0000-20.06306i$  & $0.253261i$            &         &        & $7$ & $0.0000-8.395168i$    & $0.253518i$\\
$12/27$&$1$    & $0$ &$0.0000-11.67437i$  & $3.327622i$            & $28/27$ & $1$    & $0$ & $0.0000-3.808547i$    & $\rev{0.256695i}$\\
       &       & $1$ &$0.0000-15.00199i$  & $0.253393i$            &         &        & $1$ & $0.0000-4.065242i$    & $0.256198i$\\
       &       & $2$ &$0.0000-15.25539i$  & $0.253472i$            &         &        & $2$ & $0.0000-4.321440i$    & $0.255770i$\\
       &       & $3$ &$0.0000-15.50886i$  & $0.253412i$            &         &        & $3$ & $0.0000-4.577209i$    & $0.255401i$\\
       &       & $4$ &$0.0000-15.76227i$  & $0.253412i$            &         &        & $4$ & $0.0000-4.832611i$    & $0.255085i$\\
       &       & $5$ &$0.0000-16.01568i$  & $0.253402i$            &         &        & $5$ & $0.0000-5.087695i$    & $0.254811i$\\
       &       & $6$ &$0.0000-16.26908i$  & $0.253392i$            &         &        & $6$ & $0.0000-5.342507i$    & $0.254575i$\\
       &       & $7$ &$0.0000-16.52248i$  & $0.253383i$            &         &        & $7$ & $0.0000-5.597082i$    & $0.254370i$\\
[1ex]
\hline\hline 
\end{tabular}
\end{table}

\begin{table}%[ht]
\small\setlength{\tabcolsep}{4.5pt}
\centering
\caption{Comparison of fundamental QNMs and additional overtone candidates for scalar perturbations of the nearly-extremal Hayward black hole 
 for $\ell\in\{0,1\}$, and $\gamma=1.18$. The corresponding results obtained by \cite{Zainab2024EPL} using a semi-analytical formula are included for reference. The location of the event horizon $\rho_h$ is provided in Table~\ref{table:event}. Our QNMs are computed using the Spectral Method, with reported spectral truncation order $N_{\mathrm C}=300$ and 200-digit numerical precision. Here, $\Omega$ denotes the dimensionless QNM frequency and $N$ the overtone number. Entries marked ‘N/A’ indicate unavailable data, while ‘SM’ refers to the Spectral Method.}
\label{scalarg03Zainab}
\vspace*{1em}
\begin{tabular}{||c|c|c|c|c|c|c|c|c|c|c|c|c||}
\hline\hline
$\ell$ & $N$ & $\Omega$ \cite{Zainab2024EPL}    & $\Omega$ (SM) & $\ell$ & $N$ & $\Omega$ \cite{Zainab2024EPL}    & $\Omega$ (SM) \\ [0.5ex]
\hline\hline
$0$    & $0$ & $0.112852-0.078891i$             & $0.110738-0.087901i$  & $1$    & $0$ & $0.308291-0.082590i$     & $0.307461-0.082200i$\\
       & $1$ & N/A                              & N/A                   &        & $1$ & N/A                      & $0.271661-0.255506i$\\
       & $2$ & N/A                              & N/A                   &        & $2$ & N/A                      & $0.209390-0.457941i$\\
[1ex]
\hline\hline 
\end{tabular}
\end{table}

\begin{table}%[ht]
\small\setlength{\tabcolsep}{4.5pt}
\centering
\caption{
Unclassified purely imaginary spectral candidates for scalar perturbations of the nearly-extremal Hayward black hole for $\gamma = 1.18$, and $\ell\in\{0,1\}$. The locations of the event horizon $\rho_h$ are given in Table~\ref{table:event}. The corresponding results are obtained through our spectral method, with reported spectral truncation order $N_{\mathrm C}=300$ with a precision of $200$ digits. In this context, $\Omega$ and $N$ represent the dimensionless frequency and the local row index, respectively, while $\Delta\Omega=\Omega_N-\Omega_{N+1}$. The notation 'SM' stands for Spectral Method. These entries have not been verified as isolated Green-function poles. The displayed precision is not a numerical error bound (see Secs.~\ref{sec:pole_classification} and \ref{sec:spacing_audit}).}
\label{scalarg03overdamped}
\vspace*{1em}
\begin{tabular}{||c|c|c|c|c|c|c|c|c|c|c|c|c||}
\hline\hline
$\ell$ & $N$ & $\Omega$ (SM)    & $\Delta\Omega$ & $\ell$ & $N$ & $\Omega$ (SM)     & $\Delta\Omega$ \\ [0.5ex]
\hline\hline
$0$    & $0$ &$0.000-0.570623i$         & $0.255989i$            & $1$    & $0$ & $0.000-0.115632i$         & $0.029213i$\\
       & $1$ &$0.000-0.826612i$         & $0.254606i$            &        & $1$ & $0.000-0.144844i$         & $0.029317i$\\
       & $2$ &$0.000-1.081219i$         & $0.254048i$            &        & $2$ & $0.000-0.174162i$         & $0.189467i$\\
       & $3$ &$0.000-1.335267i$         & $0.253734i$            &        & $3$ & $0.000-0.363629i$         & $0.289601i$\\
       & $4$ &$0.000-1.589000i$         & $0.253502i$            &        & $4$ & $0.000-0.653230i$         & $0.274544i$\\
       & $5$ &$0.000-1.842502i$         & $0.253307i$            &        & $5$ & $0.000-0.927774i$         & $0.269020i$\\
       & $6$ &$0.000-2.095809i$         & $0.253135i$            &        & $6$ & $0.000-1.196794i$         & $0.266397i$\\
       & $7$ &$0.000-2.348943i$         & $0.252983i$            &        & $7$ & $0.000-1.463191i$         & $0.264802i$\\
       & $8$ &$0.000-2.601926i$         & $0.252851i$            &        & $8$ & $0.000-1.727993i$         & $0.263549i$\\
       & $9$ &$0.000-2.854777i$         & $0.252737i$            &        & $9$ & $0.000-1.991542i$         & $0.262388i$\\
       &$10$ &$0.000-3.107514i$         & $0.252641i$            &        &$10$ & $0.000-2.253930i$         & $0.261262i$\\
       &$11$ &$0.000-3.360155i$         & $0.252560i$            &        &$11$ & $0.000-2.515192i$         & $0.260187i$\\
[1ex]
\hline\hline 
\end{tabular}
\end{table}

\begin{table}%[ht]
\small\setlength{\tabcolsep}{4.5pt}
\centering
\caption{Oscillatory QNM benchmarks and additional spectral candidates for electromagnetic perturbations of the non-extremal Hayward black hole for different values of $\ell$ and $\gamma$. The locations of the event horizon $\rho_h$ are given in Table~\ref{table:event}. The results are obtained using our spectral method with reported spectral truncation order $N_{\mathrm C}=300$ and 200-digit precision. Here, $\Omega$ denotes the dimensionless QNM frequency, and $N$ the overtone number. 'N/A' indicates data not available, and 'SM' refers to the Spectral Method.}
\label{emKonoplyaZeinab}
\vspace*{1em}
\begin{tabular}{||c|c|c|c|c|c|c|c|c|c|c|c|c||}
\hline\hline
$\gamma$   &$\ell$ & $N$ & $\Omega$ \cite{Konoplya2022JCAP} & $\Omega$ (SM) & $\gamma$  & $\ell$ & $N$ & $\Omega$ \cite{Zainab2024EPL} & $\Omega$ (SM) \\ [0.5ex]
\hline\hline
$0$    &$1$    & $0$ &$0.248264-0.092488i$            & $0.248263-0.092488i$          & $1.1$   & $1$    & $0$ & $0.267047-0.078538i$      & $0.266903-0.077944i$\\
       &       & $1$ &$0.214516-0.293668i$            & $0.214515-0.293668i$          &         &        & $1$ & N/A                       & $0.231752-0.241265i$\\
       &       & $2$ &$0.174774-0.525188i$            & $0.174774-0.525188i$          &         &        & $2$ & N/A                       & $0.167012-0.432085i$\\
       &       & $3$ &$0.146177-0.771909i$            & $0.146177-0.771909i$          &         &        & $3$ & N/A                       & $0.100803-0.661516i$\\
       &       & $4$ &$0.126554-1.022551i$            & $0.126554-1.022550i$          &         &        & $4$ & N/A                       & N/A\\
       &       & $5$ &$0.112253-1.273926i$            & $0.112251-1.273927i$          &         &        & $5$ & N/A                       & N/A\\
       &       & $6$ &$0.101215-1.525267i$            & $0.101225-1.525280i$          &         &        & $6$ & N/A                       & N/A\\
$1$    &$1$    & $0$ &$0.265285-0.080169i$            & $0.265285-0.080169i$          & $1.1$   & $2$    & $0$ & N/A                       & $0.483799-0.081284i$\\
       &       & $1$ &$0.233370-0.247397i$            & $0.233370-0.247397i$          &         &        & $1$ & N/A                       & $0.462853-0.246497i$\\
       &       & $2$ &$0.172918-0.437335i$            & $0.172918-0.437335i$          &         &        & $2$ & N/A                       & $0.421383-0.420352i$\\
       &       & $3$ &$0.105275-0.663349i$            & $0.105275-0.663349i$          &         &        & $3$ & N/A                       & $0.362073-0.610742i$\\
       &       & $4$ &$0.052269-0.905784i$            & $0.052288-0.905811i$          &         &        & $4$ & N/A                       & $0.293867-0.825034i$\\
       &       & $5$ &$0.015138-1.152384i$            & N/A                           &         &        & $5$ & N/A                       & $0.229922-1.062151i$\\
       &       & $6$ &$0.000-1.441i$                  & N/A                           &         &        & $6$ & N/A                       & $0.177348-1.313129i$\\
$1$    &$2$    & $0$ &N/A                             & $0.481041-0.083293i$          & $1.18$  & $1$    & $0$ & $0.268499-0.076956i$      & $0.268054-0.076020i$\\
       &       & $1$ &N/A                             & $0.461626-0.252476i$          &         &        & $1$ & N/A                       & $0.230011-0.236867i$\\
       &       & $2$ &N/A                             & $0.423772-0.429442i$          &         &        & $2$ & N/A                       & N/A\\
       &       & $3$ &N/A                             & $0.369422-0.620102i$          &         &        & $3$ & N/A                       & N/A\\
       &       & $4$ &N/A                             & $0.304495-0.831680i$          &         &        & $4$ & N/A                       & N/A\\
       &       & $5$ &N/A                             & $0.241071-1.065834i$          &         &        & $5$ & N/A                       & N/A\\
       &       & $6$ &N/A                             & $0.187625-1.314872i$          &         &        & $6$ & N/A                       & N/A\\
$1$    &$3$    & $0$ &N/A                             & $0.688489-0.084201i$          & $1.18$  & $2$    & $0$ & N/A                       & $0.486026-0.079485i$\\
       &       & $1$ &N/A                             & $0.674512-0.253936i$          &         &        & $1$ & N/A                       & $0.463321-0.241522i$\\
       &       & $2$ &N/A                             & $0.646930-0.427661i$          &         &        & $2$ & N/A                       & $0.418666-0.413643i$\\
       &       & $3$ &N/A                             & $0.606430-0.608086i$          &         &        & $3$ & N/A                       & $0.356042-0.604526i$\\
       &       & $4$ &N/A                             & $0.554096-0.798302i$          &         &        & $4$ & N/A                       & N/A\\
       &       & $5$ &N/A                             & $0.492318-1.002084i$          &         &        & $5$ & N/A                       & N/A\\
       &       & $6$ &N/A                             & $0.426227-1.222485i$          &         &        & $6$ & N/A                       & N/A\\
[1ex]
\hline\hline 
\end{tabular}
\end{table}

\begin{table}%[ht]
\small\setlength{\tabcolsep}{4.5pt}
\centering
\caption{Unclassified purely imaginary spectral candidates for electromagnetic perturbations of the non-extremal Hayward black hole for different values of $\ell$ and $\gamma$. The locations of the event horizon $\rho_h$ are given in Table~\ref{table:event}. The corresponding results are obtained through our spectral method, with reported spectral truncation order $N_{\mathrm C}=300$ with a precision of $200$ digits. In this context, $\Omega$ and $N$ represent the dimensionless frequency and the local row index, respectively, while $\Delta\Omega=\Omega_N-\Omega_{N+1}$. The notation 'SM' stands for Spectral Method. These entries have not been verified as isolated Green-function poles. The displayed precision is not a numerical error bound (see Secs.~\ref{sec:pole_classification} and \ref{sec:spacing_audit}).}
\label{em01overdamped}
\vspace*{1em}
\begin{tabular}{||c|c|c|c|c|c|c|c|c|c|c|c|c||}
\hline\hline
$\gamma$&$\ell$& $N$ &$\Omega$ (SM)        & $\Delta\Omega$         & $\gamma$& $\ell$ & $N$ & $\Omega$ (SM) & $\Delta\Omega$ \\ [0.5ex]
\hline\hline
$0$    &$1$    & $0$ &$0.0000-27.48094i$   & $0.250099i$            & $1.1$   & $1$    & $0$ & $0.0000-2.421236i$    & $0.253395i$\\
       &       & $1$ &$0.0000-27.73104i$   & $0.250188i$            &         &        & $1$ & $0.0000-2.674630i$    & $0.253241i$\\
       &       & $2$ &$0.0000-27.98123i$   & $0.249991i$            &         &        & $2$ & $0.0000-2.927872i$    & $0.253133i$\\
       &       & $3$ &$0.0000-28.23122i$   & $0.250133i$            &         &        & $3$ & $0.0000-3.181004i$    & $0.253045i$\\
       &       & $4$ &$0.0000-28.48135i$   & $0.250064i$            &         &        & $4$ & $0.0000-3.434049i$    & $0.252971i$\\
       &       & $5$ &$0.0000-28.73141i$   & $0.250080i$            &         &        & $5$ & $0.0000-3.687020i$    & $0.252908i$\\
       &       & $6$ &$0.0000-28.98150i$   & $0.250084i$            &         &        & $6$ & $0.0000-3.939928i$    & $0.252853i$\\
       &       & $7$ &$0.0000-29.23158i$   & $0.250074i$            &         &        & $7$ & $0.0000-4.192781i$    & $0.252805i$\\
$1$    &$1$    & $0$ &$0.0000-4.701594i$   & $0.252884i$            & $1.1$   & $2$    & $0$ & $0.0000-2.577889i$    & $0.259795i$\\
       &       & $1$ &$0.0000-4.954479i$   & $0.252864i$            &         &        & $1$ & $0.0000-2.837684i$    & $0.258997i$\\
       &       & $2$ &$0.0000-5.207342i$   & $0.252833i$            &         &        & $2$ & $0.0000-3.096681i$    & $0.258314i$\\
       &       & $3$ &$0.0000-5.460175i$   & $0.252806i$            &         &        & $3$ & $0.0000-3.354996i$    & $0.257721i$\\
       &       & $4$ &$0.0000-5.712981i$   & $0.252781i$            &         &        & $4$ & $0.0000-3.612716i$    & $0.257199i$\\
       &       & $5$ &$0.0000-5.965763i$   & $0.252759i$            &         &        & $5$ & $0.0000-3.869915i$    & $0.256739i$\\
       &       & $6$ &$0.0000-6.218522i$   & $\rev{0.252728i}$            &         &        & $6$ & $0.0000-4.126654i$    & $0.256331i$\\
       &       & $7$ &$0.0000-6.471250i$   & $0.252720i$            &         &        & $7$ & $0.0000-4.382985i$    & $0.255969i$\\
$1$    &$2$    & $0$ &$0.0000-4.640543i$   & $0.255857i$            & $1.18$  & $1$    & $0$ & $0.0000-0.889309i$    & $0.258291i$\\
       &       & $1$ &$0.0000-4.896400i$   & $0.255580i$            &         &        & $1$ & $0.0000-1.147600i$    & $0.256019i$\\
       &       & $2$ &$0.0000-5.151980i$   & $0.255328i$            &         &        & $2$ & $0.0000-1.403619i$    & $0.254829i$\\
       &       & $3$ &$0.0000-5.407308i$   & $0.255101i$            &         &        & $3$ & $0.0000-1.658448i$    & $0.254147i$\\
       &       & $4$ &$0.0000-5.662409i$   & $0.254898i$            &         &        & $4$ & $0.0000-1.912595i$    & $0.253728i$\\
       &       & $5$ &$0.0000-5.917307i$   & $0.254715i$            &         &        & $5$ & $0.0000-2.166323i$    & $0.253452i$\\
       &       & $6$ &$0.0000-6.172022i$   & $0.254551i$            &         &        & $6$ & $0.0000-2.419775i$    & $0.253259i$\\
       &       & $7$ &$0.0000-6.426573i$   & $0.254402i$            &         &        & $7$ & $0.0000-2.673034i$    & $0.253115i$\\
$1$    &$3$    & $0$ &$0.0000-5.062348i$   & $0.259749i$            & $1.18$  & $2$    & $0$ & $0.0000-2.577514i$    & $0.259718i$\\
       &       & $1$ &$0.0000-5.322096i$   & $0.259174i$            &         &        & $1$ & $0.0000-2.837231i$    & $0.258897i$\\
       &       & $2$ &$0.0000-5.581270i$   & $0.258647i$            &         &        & $2$ & $0.0000-3.096128i$    & $0.258194i$\\
       &       & $3$ &$0.0000-5.839916i$   & $0.258166i$            &         &        & $3$ & $0.0000-3.354322i$    & $0.257583i$\\
       &       & $4$ &$0.0000-6.098082i$   & $0.257730i$            &         &        & $4$ & $0.0000-3.611905i$    & $0.257047i$\\
       &       & $5$ &$0.0000-6.355812i$   & $0.257335i$            &         &        & $5$ & $0.0000-3.868952i$    & $0.256576i$\\
       &       & $6$ &$0.0000-6.613146i$   & $0.256977i$            &         &        & $6$ & $0.0000-4.125527i$    & $0.256160i$\\
       &       & $7$ &$0.0000-6.870123i$   & $0.256654i$            &         &        & $7$ & $0.0000-4.381687i$    & $0.255792i$\\
[1ex]
\hline\hline 
\end{tabular}
\end{table}

\begin{table}%[ht]
\small\setlength{\tabcolsep}{4.5pt}
\centering
\caption{Oscillatory QNM benchmarks and additional spectral candidates for the adopted axial effective model of the non-extremal Hayward black hole for different values of $\ell$ and $\gamma$. The locations of the event horizon $\rho_h$ are given in Table~\ref{table:event}. The results are obtained using our spectral method with reported spectral truncation order $N_{\mathrm C}=200$ and 200-digit precision. Here, $\Omega$ denotes the dimensionless QNM frequency, and $N$ the overtone number. 'N/A' indicates data not available, and 'SM' refers to the Spectral Method. The potential is $U_2^{\rm here}$ in Eq.~\eqref{eq:axial_here}, which differs from the effective-source closure of Refs.~\cite{Malik2025IJTP,Bolokhov2026EPJC}.}
\label{tensor01}
\vspace*{1em}
\begin{tabular}{||c|c|c|c|c|c|c|c|c|c|c|c|c||}
\hline\hline
$\gamma$ &$\ell$ & $N$ & $\Omega$ (SM) & $\gamma$     & $\ell$ & $N$ & $\Omega$ (SM) \\ [0.5ex]
\hline\hline
$0$      &$2$    & $0$ & $0.373672-0.088962i$           & $1$          & $2$    & $0$ & $0.402067-0.074637i$\\
         &       & $1$ & $0.346711-0.273915i$           &              &        & $1$ & $0.384948-0.227230i$\\
         &       & $2$ & $0.301054-0.478277i$           &              &        & $2$ & $0.351849-0.389555i$\\
         &       & $3$ & $0.251505-0.705148i$           &              &        & $3$ & $0.303852-0.567976i$\\
         &       & $4$ & $0.207515-0.946845i$           &              &        & $4$ & $0.245048-0.771395i$\\
         &       & $5$ & $0.169299-1.195608i$           &              &        & $5$ & $0.189349-1.002833i$\\
         &       & $6$ & $0.133253-1.447910i$           &              &        & $6$ & $0.146653-1.252056i$\\
$0.5$    &$2$    & $0$ & $0.386120-0.083547i$           & $1$          & $3$    & $0$ & $0.634333-0.080456i$\\
         &       & $1$ & $0.365774-0.256153i$           &              &        & $1$ & $0.620639-0.242789i$\\
         &       & $2$ & $0.331612-0.443718i$           &              &        & $2$ & $0.593669-0.409356i$\\
         &       & $3$ & $0.294565-0.649154i$           &              &        & $3$ & $0.554143-0.582933i$\\
         &       & $4$ & $0.262226-0.867843i$           &              &        & $4$ & $0.502992-0.766632i$\\
         &       & $5$ & $0.235726-1.093994i$           &              &        & $5$ & $0.442241-0.964554i$\\
         &       & $6$ & $0.213178-1.324082i$           &              &        & $6$ & $0.377115-1.180511i$\\
$0.5$    &$3$    & $0$ & $0.614939-0.088052i$           & $1.18$       & $2$    & $0$ & $0.408845-0.069493i$\\
         &       & $1$ & $0.600863-0.266607i$           &              &        & $1$ & $0.388809-0.212218i$\\
         &       & $2$ & $0.574702-0.452164i$           &              &        & $2$ & $0.348527-0.367404i$\\
         &       & $3$ & $0.540332-0.647932i$           &              &        & $3$ & $0.290649-0.546172i$\\
         &       & $4$ & $0.502491-0.854668i$           &              &        & $4$ & N/A\\
         &       & $5$ & $0.464967-1.070729i$           &              &        & $5$ & N/A\\
         &       & $6$ & $0.429573-1.293480i$           &              &        & $6$ & N/A\\
$0.5$    &$4$    & $0$ & $0.828150-0.089688i$           & $1.18$       & $3$    & $0$ & $0.642615-0.076211i$\\
         &       & $1$ & $0.817332-0.270470i$           &              &        & $1$ & $0.626637-0.230230i$\\
         &       & $2$ & $0.796579-0.455354i$           &              &        & $2$ & $0.594738-0.389316i$\\
         &       & $3$ & $0.767653-0.646609i$           &              &        & $3$ & $0.547497-0.557721i$\\
         &       & $4$ & $0.733025-0.845669i$           &              &        & $4$ & $0.487087-0.740539i$\\
         &       & $5$ & $0.695389-1.052817i$           &              &        & $5$ & N/A\\
         &       & $6$ & $0.657024-1.267259i$           &              &        & $6$ & N/A\\
[1ex]
\hline\hline 
\end{tabular}
\end{table}

\begin{table}%[ht]
\small\setlength{\tabcolsep}{4.5pt}
\centering
\caption{Unclassified purely imaginary spectral candidates for the adopted axial effective model of the non-extremal Hayward black hole for different values of $\ell$ and $\gamma$. The locations of the event horizon $\rho_h$ are given in Table~\ref{table:event}. The corresponding results are obtained through our spectral method, with reported spectral truncation order $N_{\mathrm C}=200$ with a precision of $200$ digits. In this context, $\Omega$ and $N$ represent the dimensionless frequency and the local row index, respectively, while $\Delta\Omega=\Omega_N-\Omega_{N+1}$. The notation 'SM' stands for Spectral Method. These entries have not been verified as isolated Green-function poles. The displayed precision is not a numerical error bound (see Secs.~\ref{sec:pole_classification} and \ref{sec:spacing_audit}).}
\label{tensor01overdamped}
\vspace*{1em}
\begin{tabular}{||c|c|c|c|c|c|c|c|c|c|c|c|c||}
\hline\hline
$\gamma$  &$\ell$ & $N$ &$\Omega$ (SM) & $\Delta\Omega$ & $\gamma$     & $\ell$ & $N$ & $\Omega$ (SM) & $\Delta\Omega$ \\ [0.5ex]
\hline\hline
$0$       &$2$    & $0$ &$0.0000-27.840510i$    & $0.250245i$            & $1$          & $2$    & $0$ & $0.0000-4.339753i$    & $0.255175i$\\
          &       & $1$ &$0.0000-28.090755i$    & $0.250158i$            &              &        & $1$ & $0.0000-4.594928i$    & $0.254891i$\\
          &       & $2$ &$0.0000-28.340913i$    & $0.250117i$            &              &        & $2$ & $0.0000-4.849819i$    & $0.254644i$\\
          &       & $3$ &$0.0000-28.591030i$    & $0.250172i$            &              &        & $3$ & $0.0000-5.104463i$    & $0.254430i$\\
          &       & $4$ &$0.0000-28.841203i$    & $0.250133i$            &              &        & $4$ & $0.0000-5.358893i$    & $0.254246i$\\
          &       & $5$ &$0.0000-29.091335i$    & $0.250147i$            &              &        & $5$ & $0.0000-5.613139i$    & $0.254085i$\\
          &       & $6$ &$0.0000-29.341482i$    & $0.250142i$            &              &        & $6$ & $0.0000-5.867224i$    & $0.253945i$\\
          &       & $7$ &$0.0000-29.591624i$    & $0.250138i$            &              &        & $7$ & $0.0000-6.121169i$    & $0.253821i$\\
$0.5$     &$2$    & $0$ &$0.0000-13.746426i$    & $0.253426i$            & $1$          & $3$    & $0$ & $0.0000-4.513917i$    & $0.259929i$\\
          &       & $1$ &$0.0000-13.999852i$    & $0.253410i$            &              &        & $1$ & $0.0000-4.773846i$    & $0.259096i$\\
          &       & $2$ &$0.0000-14.253261i$    & $0.253385i$            &              &        & $2$ & $0.0000-5.032943i$    & $0.258419i$\\
          &       & $3$ &$0.0000-14.506647i$    & $0.253376i$            &              &        & $3$ & $0.0000-5.291362i$    & $0.257827i$\\
          &       & $4$ &$0.0000-14.760022i$    & $0.253367i$            &              &        & $4$ & $0.0000-5.549189i$    & $0.257311i$\\
          &       & $5$ &$0.0000-15.013389i$    & $0.253356i$            &              &        & $5$ & $0.0000-5.806500i$    & $0.256861i$\\
          &       & $6$ &$0.0000-15.266745i$    & $0.253347i$            &              &        & $6$ & $0.0000-6.063361i$    & $0.256468i$\\
          &       & $7$ &$0.0000-15.520092i$    & $0.253338i$            &              &        & $7$ & $0.0000-6.319828i$    & $0.256122i$\\
$0.5$     &$3$    & $0$ &$0.0000-13.715917i$    & $0.254052i$            & $1.18$       & $2$    & $0$ & $0.0000-2.538970i$    & $0.258992i$\\
          &       & $1$ &$0.0000-13.969969i$    & $0.253952i$            &              &        & $1$ & $0.0000-2.797962i$    & $0.257996i$\\
          &       & $2$ &$0.0000-14.223921i$    & $0.253920i$            &              &        & $2$ & $0.0000-3.055958i$    & $0.257170i$\\
          &       & $3$ &$0.0000-14.477841i$    & $0.253890i$            &              &        & $3$ & $0.0000-3.313128i$    & $0.256487i$\\
          &       & $4$ &$0.0000-14.731731i$    & $0.253862i$            &              &        & $4$ & $0.0000-3.569615i$    & $0.255923i$\\
          &       & $5$ &$0.0000-14.985593i$    & $0.253834i$            &              &        & $5$ & $0.0000-3.825537i$    & $0.255453i$\\
          &       & $6$ &$0.0000-15.239427i$    & $0.253808i$            &              &        & $6$ & $0.0000-4.080991i$    & $0.255061i$\\
          &       & $7$ &$0.0000-15.493234i$    & $0.253783i$            &              &        & $7$ & $0.0000-4.336052i$    & $0.254732i$\\
$0.5$     &$4$    & $0$ &$0.0000-13.926883i$    & $0.254837i$            & $1.18$       & $3$    & $0$ & $0.0000-4.514662i$    & $0.259184i$\\
          &       & $1$ &$0.0000-14.181720i$    & $0.254774i$            &              &        & $1$ & $0.0000-4.773846i$    & $0.258417i$\\
          &       & $2$ &$0.0000-14.436494i$    & $0.254708i$            &              &        & $2$ & $0.0000-5.032263i$    & $0.257756i$\\
          &       & $3$ &$0.0000-14.691202i$    & $0.254646i$            &              &        & $3$ & $0.0000-5.290019i$    & $0.257186i$\\
          &       & $4$ &$0.0000-14.945848i$    & $0.254587i$            &              &        & $4$ & $0.0000-5.547205i$    & $0.256693i$\\
          &       & $5$ &$0.0000-15.200434i$    & $0.254531i$            &              &        & $5$ & $0.0000-5.803898i$    & $0.256266i$\\
          &       & $6$ &$0.0000-15.454965i$    & $0.254479i$            &              &        & $6$ & $0.0000-6.060163i$    & $0.255894i$\\
          &       & $7$ &$0.0000-15.709444i$    & $0.254429i$            &              &        & $7$ & $0.0000-6.316058i$    & $0.255569i$\\
[1ex]
\hline\hline 
\end{tabular}
\end{table}

\begin{table}%[ht]
\small\setlength{\tabcolsep}{4.5pt}
\centering
\caption{Oscillatory QNM benchmarks and additional spectral candidates for scalar perturbations of the extremal Hayward black hole across various angular momentum values $\ell$. The QNMs are computed using the Spectral Method with reported spectral truncation order $N_{\mathrm C}=200$ and 200-digit numerical precision. Here, $\Omega$ denotes the dimensionless QNM frequency and $N$ the overtone number. Entries marked ‘N/A’ indicate unavailable data, while ‘SM’ denotes results obtained via the Spectral Method. For comparison, \cite{Konoplya2022JCAP} reports only the fundamental mode for $\ell = 0$, computed using the WKB approximation and the time-domain technique, yielding $\Omega_{\text{WKB}} = 0.110844 - 0.087919i$ and $\Omega_{\text{TD}} = 0.110664 - 0.087749i$, respectively.}
\label{scalargextreme}
\vspace*{1em}
\begin{tabular}{||c|c|c|c|c|c|c|c|c|c|c|c|c||}
\hline\hline
$\ell$ & $N$ & $\Omega$ (SM)   & $\ell$ & $N$ & $\Omega$ (SM) \\ [0.5ex]
\hline\hline
$0$    & $0$ &$0.110697-0.087879i$              & $3$    & $0$ & $0.712751-0.081531i$\\
       & $1$ &$0.061620-0.312487i$              &        & $1$ & $0.696506-0.246168i$\\
       & $2$ &N/A                               &        & $2$ & $0.664233-0.415751i$\\
       & $3$ &N/A                               &        & $3$ & $0.616791-0.594232i$\\
       & $4$ &N/A                               &        & $4$ & $0.556554-0.786083i$\\
       & $5$ &N/A                               &        & $5$ & $0.488362-0.995170i$\\
       & $6$ &N/A                               &        & $6$ & $0.419102-1.222557i$\\
       & $7$ &N/A                               &        & $7$ & $0.355007-1.465380i$\\
       & $8$ &N/A                               &        & $8$ & $0.299194-1.718697i$\\
$1$    & $0$ &$0.307507-0.082099i$              & $4$    & $0$ & $0.915909-0.081497i$\\
       & $1$ &$0.271554-0.255290i$              &        & $1$ & $0.903212-0.245436i$\\
       & $2$ &$0.209147-0.457774i$              &        & $2$ & $0.877894-0.412296i$\\
       & $3$ &$0.147781-0.696550i$              &        & $3$ & $0.840226-0.584289i$\\
       & $4$ &$0.103545-0.953106i$              &        & $4$ & $0.790920-0.763952i$\\
       & $5$ &N/A                               &        & $5$ & $0.731512-0.954051i$\\
       & $6$ &N/A                               &        & $6$ & $0.664747-1.157122i$\\
       & $7$ &N/A                               &        & $7$ & $0.594572-1.374617i$\\
       & $8$ &N/A                               &        & $8$ & $0.525356-1.606115i$\\
$2$    & $0$ &$0.509795-0.081632i$              & $5$    & $0$ & $1.119153-0.081482i$\\
       & $1$ &$0.487305-0.248027i$              &        & $1$ & $1.108737-0.245075i$\\
       & $2$ &$0.443288-0.424580i$              &        & $2$ & $1.087939-0.410597i$\\
       & $3$ &$0.381924-0.619622i$              &        & $3$ & $1.056874-0.579458i$\\
       & $4$ &$0.313437-0.839255i$              &        & $4$ & $1.015817-0.753235i$\\
       & $5$ &$0.250339-1.080927i$              &        & $5$ & $0.965352-0.933688i$\\
       & $6$ &$0.198638-1.335717i$              &        & $6$ & $0.906555-1.122689i$\\
       & $7$ &$0.157684-1.596319i$              &        & $7$ & $0.841179-1.322002i$\\
       & $8$ &N/A                               &        & $8$ & $0.771689-1.532893i$\\
       [1ex]
\hline\hline 
\end{tabular}
\end{table}

\begin{table}%[ht]
\small\setlength{\tabcolsep}{4.5pt}
\centering
\caption{Unclassified purely imaginary spectral candidates for scalar perturbations of the extremal Hayward black hole for several values of the angular momentum $\ell$. The corresponding results are obtained through our spectral method, with reported spectral truncation order $N_{\mathrm C}=200$ with a precision of $200$ digits. In this context, $\Omega$ and $N$ represent the dimensionless frequency and the local row index, respectively, while $\Delta\Omega=\Omega_N-\Omega_{N+1}$. The notation 'SM' stands for Spectral Method. These entries have not been verified as isolated Green-function poles. The displayed precision is not a numerical error bound (see Secs.~\ref{sec:pole_classification} and \ref{sec:spacing_audit}).}
\label{scalargextremeoverdamped}
\vspace*{1em}
\begin{tabular}{||c|c|c|c|c|c|c|c|c|c|c|c|c||}
\hline\hline
$\ell$ & $N$ & $\Omega$ (SM) & $\Delta\Omega$  & $\ell$ & $N$ & $\Omega$ (SM) & $\Delta\Omega$ \\ [0.5ex]
\hline\hline
$0$    & $0$ &$0.0000-14.416315i$    & $0.252022i$              & $3$    & $0$ & $0.0000-13.074824i$    & $0.253603i$\\
       & $1$ &$0.0000-14.668337i$    & $0.252106i$              &        & $1$ & $0.0000-13.328427i$    & $0.253287i$\\
       & $2$ &$0.0000-14.920443i$    & $0.252013i$              &        & $2$ & $0.0000-13.581714i$    & $0.253399i$\\
       & $3$ &$0.0000-15.172456i$    & $0.252107i$              &        & $3$ & $0.0000-13.835114i$    & $0.253303i$\\
       & $4$ &$0.0000-15.424563i$    & $0.252026i$              &        & $4$ & $0.0000-14.088417i$    & $0.253248i$\\
       & $5$ &$0.0000-15.676590i$    & $0.252092i$              &        & $5$ & $0.0000-14.341664i$    & $0.253271i$\\
       & $6$ &$0.0000-15.928682i$    & $0.252050i$              &        & $6$ & $0.0000-14.594935i$    & $0.253151i$\\
       & $7$ &$0.0000-16.180732i$    & $0.252069i$              &        & $7$ & $0.0000-14.848086i$    & $0.253213i$\\
$1$    & $0$ &$0.0000-14.655532i$    & $0.252282i$              & $4$    & $0$ & $0.0000-12.262121i$    & $0.254486i$\\
       & $1$ &$0.0000-14.907814i$    & $0.252241i$              &        & $1$ & $0.0000-12.516607i$    & $0.254461i$\\
       & $2$ &$0.0000-15.160055i$    & $0.252276i$              &        & $2$ & $0.0000-12.771069i$    & $0.254219i$\\
       & $3$ &$0.0000-15.412331i$    & $0.252236i$              &        & $3$ & $0.0000-13.025288i$    & $0.254379i$\\
       & $4$ &$0.0000-15.664567i$    & $0.252254i$              &        & $4$ & $0.0000-13.279667i$    & $0.254046i$\\
       & $5$ &$0.0000-15.916821i$    & $0.252245i$              &        & $5$ & $0.0000-13.533714i$    & $0.254210i$\\
       & $6$ &$0.0000-16.169066i$    & $0.252224i$              &        & $6$ & $0.0000-13.787924i$    & $0.253968i$\\
       & $7$ &$0.0000-16.421289i$    & $0.252261i$              &        & $7$ & $0.0000-14.041892i$    & $0.254017i$\\
$2$    & $0$ &$0.0000-14.630639i$    & $0.252628i$              & $5$    & $0$ & $0.0000-10.149512i$    & $0.256866i$\\
       & $1$ &$0.0000-14.883267i$    & $0.252665i$              &        & $1$ & $0.0000-10.406379i$    & $0.256883i$\\
       & $2$ &$0.0000-15.135931i$    & $0.252601i$              &        & $2$ & $0.0000-10.663262i$    & $0.256527i$\\
       & $3$ &$0.0000-15.388532i$    & $0.252631i$              &        & $3$ & $0.0000-10.919789i$    & $0.256376i$\\
       & $4$ &$0.0000-15.641163i$    & $0.252564i$              &        & $4$ & $0.0000-11.176165i$    & $0.256148i$\\
       & $5$ &$0.0000-15.893728i$    & $0.252612i$              &        & $5$ & $0.0000-11.432313i$    & $0.256140i$\\ 
       & $6$ &$0.0000-16.146340i$    & $0.252522i$              &        & $6$ & $0.0000-11.688453i$    & $0.255765i$\\
       & $7$ &$0.0000-16.398861i$    & $0.252602i$              &        & $7$ & $0.0000-11.944218i$    & $0.255799i$\\
       [1ex]
\hline\hline 
\end{tabular}
\end{table}

\begin{table}%[ht]
\small\setlength{\tabcolsep}{4.5pt}
\centering
\caption{Oscillatory QNM benchmarks and additional spectral candidates for electromagnetic perturbations of the extremal Hayward black hole for various angular momentum values $\ell$. Our QNMs are computed using the Spectral Method, with reported spectral truncation order $N_{\mathrm C}=200$ and 200-digit numerical precision. Here, $\Omega$ denotes the dimensionless QNM frequency and $N$ the overtone number. Entries marked ‘N/A’ indicate unavailable data, while ‘SM’ refers to the Spectral Method.}
\label{emextreme}
\vspace*{1em}
\begin{tabular}{||c|c|c|c|c|c|c|c|c|c|c|c|c||}
\hline\hline
$\ell$ & $N$ & $\Omega$ (SM)   & $\ell$ & $N$ & $\Omega$ (SM) \\ [0.5ex]
\hline\hline
$1$    & $0$ &$0.268124-0.075892i$              & $4$    & $0$ & $0.902769-0.080788i$\\
       & $1$ &$0.229887-0.236597i$              &        & $1$ & $0.890024-0.243308i$\\
       & $2$ &$0.162024-0.428453i$              &        & $2$ & $0.864603-0.408746i$\\
       & $3$ &$0.095517-0.660214i$              &        & $3$ & $0.826760-0.579322i$\\
       & $4$ &N/A                               &        & $4$ & $0.777186-0.757598i$\\
       & $5$ &N/A                               &        & $5$ & $0.717411-0.946384i$\\
       & $6$ &N/A                               &        & $6$ & $0.650203-1.148277i$\\
       & $7$ &N/A                               &        & $7$ & $0.579581-1.364772i$\\
       & $8$ &N/A                               &        & $8$ & $0.510004-1.595438i$\\
$2$    & $0$ &$0.486170-0.079362i$              & $5$    & $0$ & $1.108399-0.081006i$\\
       & $1$ &$0.463334-0.241196i$              &        & $1$ & $1.097958-0.243648i$\\
       & $2$ &$0.418470-0.413229i$              &        & $2$ & $1.077108-0.408217i$\\
       & $3$ &$0.355653-0.604157i$              &        & $3$ & $1.045956-0.576125i$\\
       & $4$ &$0.285539-0.820555i$              &        & $4$ & $1.004769-0.748956i$\\
       & $5$ &$0.221393-1.059618i$              &        & $5$ & $0.954122-0.928482i$\\
       & $6$ &$0.169281-1.311734i$              &        & $6$ & $0.895090-1.116599i$\\
       & $7$ &$0.128217-1.569304i$              &        & $7$ & $0.829433-1.315096i$\\
       & $8$ &N/A                               &        & $8$ & $0.759642-1.525265i$\\
$3$    & $0$ &$0.695864-0.080364i$              & $6$    & $0$ & $1.313340-0.081133i$\\
       & $1$ &$0.679510-0.242661i$              &        & $1$ & $1.304497-0.243848i$\\
       & $2$ &$0.646990-0.409902i$              &        & $2$ & $1.286831-0.407933i$\\
       & $3$ &$0.599112-0.586082i$              &        & $3$ & $1.260393-0.574364i$\\
       & $4$ &$0.538224-0.775786i$              &        & $4$ & $1.225309-0.744213i$\\
       & $5$ &$0.469255-0.983036i$              &        & $5$ & $1.181838-0.918666i$\\
       & $6$ &$0.399313-1.208947i$              &        & $6$ & $1.130456-1.099021i$\\
       & $7$ &$0.334824-1.450512i$              &        & $7$ & $1.071958-1.286637i$\\
       & $8$ &$0.278926-1.702592i$              &        & $8$ & $1.007560-1.482816i$\\
       [1ex]
\hline\hline 
\end{tabular}
\end{table}

\begin{table}%[ht]
\small\setlength{\tabcolsep}{4.5pt}
\centering
\caption{Unclassified purely imaginary spectral candidates for electromagnetic perturbations of the extremal Hayward black hole for several values of the angular momentum $\ell$. The corresponding results are obtained through our spectral method, with reported spectral truncation order $N_{\mathrm C}=200$ with a precision of $200$ digits. In this context, $\Omega$ and $N$ represent the dimensionless frequency and the local row index, respectively, while $\Delta\Omega=\Omega_N-\Omega_{N+1}$. The notation 'SM' stands for Spectral Method. These entries have not been verified as isolated Green-function poles. The displayed precision is not a numerical error bound (see Secs.~\ref{sec:pole_classification} and \ref{sec:spacing_audit}).}
\label{emextremeoverdamped}
\vspace*{1em}
\begin{tabular}{||c|c|c|c|c|c|c|c|c|c|c|c|c||}
\hline\hline
$\ell$ & $N$ & $\Omega$ (SM) & $\Delta\Omega$  & $\ell$ & $N$ & $\Omega$ (SM) & $\Delta\Omega$ \\ [0.5ex]
\hline\hline
$1$    & $0$ &$0.0000-0.000423i$    & $15.835615i$              & $4$    & $0$ & $0.0000-15.511275i$    & $0.253324i$\\
       & $1$ &$0.0000-15.836038i$   & $1.766831i$               &        & $1$ & $0.0000-15.764599i$    & $0.253282i$\\
       & $2$ &$0.0000-17.602869i$   & N/A                       &        & $2$ & $0.0000-16.017881i$    & $0.253227i$\\
       & $3$ &N/A                   & N/A                       &        & $3$ & $0.0000-16.271108i$    & $0.253222i$\\
       & $4$ &N/A                   & N/A                       &        & $4$ & $0.0000-16.524330i$    & $0.253162i$\\
       & $5$ &N/A                   & N/A                       &        & $5$ & $0.0000-16.777492i$    & $0.253148i$\\
       & $6$ &N/A                   & N/A                       &        & $6$ & $0.0000-17.030640i$    & $0.253119i$\\
       & $7$ &N/A                   & N/A                       &        & $7$ & $0.0000-17.283759i$    & $0.253070i$\\
$2$    & $0$ &$0.0000-15.064141i$   & $0.252074i$               & $5$    & $0$ & $0.0000-13.435263i$    & $0.254631i$\\
       & $1$ &$0.0000-15.316215i$   & $0.252844i$               &        & $1$ & $0.0000-13.689893i$    & $0.254402i$\\
       & $2$ &$0.0000-15.569059i$   & $0.251948i$               &        & $2$ & $0.0000-13.944296i$    & $0.254514i$\\
       & $3$ &$0.0000-15.821007i$   & $1.514381i$               &        & $3$ & $0.0000-14.198810i$    & $0.254210i$\\
       & $4$ &$0.0000-17.335388i$   & $0.252611i$               &        & $4$ & $0.0000-14.453020i$    & $0.254360i$\\
       & $5$ &$0.0000-17.587999i$   & $0.252195i$               &        & $5$ & $0.0000-14.707380i$    & $0.254080i$\\
       & $6$ &$0.0000-17.840193i$   & $0.252395i$               &        & $6$ & $0.0000-14.961460i$    & $0.254192i$\\
       & $7$ &$0.0000-18.092588i$   & $0.252363i$               &        & $7$ & $0.0000-15.215652i$    & $0.253977i$\\
$3$    & $0$ &$0.0000-16.555398i$   & $0.252575i$               & $6$    & $0$ & $0.0000-10.552510i$    & $0.257703i$\\
       & $1$ &$0.0000-16.807972i$   & $0.252815i$               &        & $1$ & $0.0000-10.810213i$    & $0.257736i$\\
       & $2$ &$0.0000-17.060787i$   & $0.252596i$               &        & $2$ & $0.0000-11.067949i$    & $0.257310i$\\
       & $3$ &$0.0000-17.313383i$   & $0.252713i$               &        & $3$ & $0.0000-11.325259i$    & $0.257117i$\\
       & $4$ &$0.0000-17.566096i$   & $0.252638i$               &        & $4$ & $0.0000-11.582377i$    & $0.256863i$\\
       & $5$ &$0.0000-17.818734i$   & $0.252616i$               &        & $5$ & $0.0000-11.839240i$    & $0.256786i$\\   
       & $6$ &$0.0000-18.071350i$   & $0.252654i$               &        & $6$ & $0.0000-12.096025i$    & $0.256388i$\\
       & $7$ &$0.0000-18.324004i$   & $0.252580i$               &        & $7$ & $0.0000-12.352414i$    & $0.256416i$\\
       [1ex]
\hline\hline 
\end{tabular}
\end{table}

\begin{table}%[ht]
\small\setlength{\tabcolsep}{4.5pt}
\centering
\caption{Oscillatory QNM benchmarks and additional spectral candidates for the adopted axial effective model of the extremal Hayward black hole for various angular momentum values $\ell$. Our QNMs are computed using the Spectral Method, with reported spectral truncation order $N_{\mathrm C}=200$ and 200-digit numerical precision. Here, $\Omega$ denotes the dimensionless QNM frequency and $N$ the overtone number. Entries marked ‘N/A’ indicate unavailable data, while ‘SM’ refers to the Spectral Method. The potential is $U_2^{\rm here}$ in Eq.~\eqref{eq:axial_here}, which differs from the effective-source closure of Refs.~\cite{Malik2025IJTP,Bolokhov2026EPJC}.}
\label{tensorextreme}
\vspace*{1em}
\begin{tabular}{||c|c|c|c|c|c|c|c|c|c|c|c|c||}
\hline\hline
$\ell$ & $N$ & $\Omega$ (SM)   & $\ell$ & $N$ & $\Omega$ (SM) \\ [0.5ex]
\hline\hline
$2$    & $0$ &$0.409045-0.069319i$              & $5$    & $0$ & $1.075531-0.079455i$\\
       & $1$ &$0.388860-0.211760i$              &        & $1$ & $1.065072-0.238995i$\\
       & $2$ &$0.348340-0.366829i$              &        & $2$ & $1.044178-0.400470i$\\
       & $3$ &$0.290252-0.545675i$              &        & $3$ & $1.012932-0.565303i$\\
       & $4$ &$0.226046-0.756693i$              &        & $4$ & $0.971577-0.735105i$\\
       & $5$ &$0.171039-0.994994i$              &        & $5$ & $0.920661-0.911700i$\\
       & $6$ &$0.130534-1.247506i$              &        & $6$ & $0.861246-1.097057i$\\
       & $7$ &N/A                               &        & $7$ & $0.795117-1.293061i$\\
       & $8$ &N/A                               &        & $8$ & $0.724846-1.501090i$\\
$3$    & $0$ &$0.642865-0.076070i$              & $6$    & $0$ & $1.285669-0.080047i$\\
       & $1$ &$0.626779-0.229831i$              &        & $1$ & $1.276807-0.240590i$\\
       & $2$ &$0.594689-0.388733i$              &        & $2$ & $1.259097-0.402504i$\\
       & $3$ &$0.547214-0.557060i$              &        & $3$ & $1.232584-0.566770i$\\
       & $4$ &$0.486585-0.739916i$              &        & $4$ & $1.197379-0.734468i$\\
       & $5$ &$0.417977-0.941966i$              &        & $5$ & $1.153727-0.906801i$\\
       & $6$ &$0.349196-1.164575i$              &        & $6$ & $1.102091-1.085094i$\\
       & $7$ &$0.287232-1.404278i$              &        & $7$ & $1.043263-1.270743i$\\
       & $8$ &$0.235199-1.655274i$              &        & $8$ & $0.978467-1.465099i$\\
$4$    & $0$ &$0.862242-0.078379i$              & $7$    & $0$ & $1.493961-0.080409i$\\
       & $1$ &$0.849515-0.236089i$              &        & $1$ & $1.486277-0.241563i$\\
       & $2$ &$0.824101-0.396748i$              &        & $2$ & $1.470918-0.403740i$\\
       & $3$ &$0.786200-0.562628i$              &        & $3$ & $1.447910-0.567661i$\\
       & $4$ &$0.736439-0.736395i$              &        & $4$ & $1.417313-0.734101i$\\
       & $5$ &$0.676319-0.921035i$              &        & $5$ & $1.379254-0.903909i$\\
       & $6$ &$0.608677-1.119364i$              &        & $6$ & $1.333961-1.078016i$\\
       & $7$ &$0.537751-1.333048i$              &        & $7$ & $1.281825-1.257427i$\\
       & $8$ &$0.468281-1.561644i$              &        & $8$ & $1.223458-1.443186i$\\
       [1ex]
\hline\hline 
\end{tabular}
\end{table}

\begin{table}%[ht]
\small\setlength{\tabcolsep}{4.5pt}
\centering
\caption{Unclassified purely imaginary spectral candidates for the adopted axial effective model of the extremal Hayward black hole for several values of the angular momentum $\ell$. The corresponding results are obtained through our spectral method, with reported spectral truncation order $N_{\mathrm C}=200$ with a precision of $200$ digits. In this context, $\Omega$ and $N$ represent the dimensionless frequency and the local row index, respectively, while $\Delta\Omega=\Omega_N-\Omega_{N+1}$. The notation 'SM' stands for Spectral Method. These entries have not been verified as isolated Green-function poles. The displayed precision is not a numerical error bound (see Secs.~\ref{sec:pole_classification} and \ref{sec:spacing_audit}).}
\label{tensorextremeoverdamped}
\vspace*{1em}
\begin{tabular}{||c|c|c|c|c|c|c|c|c|c|c|c|c||}
\hline\hline
$\ell$ & $N$ & $\Omega$ (SM) & $\Delta\Omega$  & $\ell$ & $N$ & $\Omega$ (SM) & $\Delta\Omega$ \\ [0.5ex]
\hline\hline
$2$    & $0$ &$0.0000-10.390118i$    & $0.252361i$              & $5$    & $0$ & $0.0000-9.774290i$     & $0.255112i$\\
       & $1$ &$0.0000-10.642480i$    & $0.252236i$              &        & $1$ & $0.0000-10.029403i$    & $0.255010i$\\
       & $2$ &$0.0000-10.894716i$    & $0.252350i$              &        & $2$ & $0.0000-10.284412i$    & $0.254835i$\\
       & $3$ &$0.0000-11.147066i$    & $0.252317i$              &        & $3$ & $0.0000-10.539247i$    & $0.254679i$\\
       & $4$ &$0.0000-11.399383i$    & $0.252266i$              &        & $4$ & $0.0000-10.793927i$    & $0.254533i$\\
       & $5$ &$0.0000-11.651649i$    & $0.252245i$              &        & $5$ & $0.0000-11.048460i$    & $0.254455i$\\
       & $6$ &$0.0000-11.903894i$    & $0.252345i$              &        & $6$ & $0.0000-11.302914i$    & $0.254318i$\\
       & $7$ &$0.0000-12.156239i$    & $0.252164i$              &        & $7$ & $0.0000-11.557233i$    & $0.254222i$\\
$3$    & $0$ &$0.0000-7.578863i$     & $0.253617i$              & $6$    & $0$ & $0.0000-10.482250i$    & $0.256299i$\\
       & $1$ &$0.0000-7.832480i$     & $0.253397i$              &        & $1$ & $0.0000-10.738549i$    & $0.256135i$\\
       & $2$ &$0.0000-8.085876i$     & $0.506707i$              &        & $2$ & $0.0000-10.994684i$    & $0.255902i$\\
       & $3$ &$0.0000-8.592583i$     & $0.253371i$              &        & $3$ & $0.0000-11.250586i$    & $0.255635i$\\
       & $4$ &$0.0000-8.845954i$     & $0.253152i$              &        & $4$ & $0.0000-11.506220i$    & $0.255551i$\\
       & $5$ &$0.0000-9.099107i$     & $0.253219i$              &        & $5$ & $0.0000-11.761771i$    & $0.255335i$\\
       & $6$ &$0.0000-9.352325i$     & $0.252980i$              &        & $6$ & $0.0000-12.017106i$    & $0.255187i$\\
       & $7$ &$0.0000-9.605306i$     & $0.253019i$              &        & $7$ & $0.0000-12.272293i$    & $0.255048i$\\
$4$    & $0$ &$0.0000-8.805544i$     & $0.254241i$              & $7$    & $0$ & $0.0000-10.925305i$    & $0.257943i$\\
       & $1$ &$0.0000-9.059785i$     & $0.254162i$              &        & $1$ & $0.0000-11.183248i$    & $0.257703i$\\
       & $2$ &$0.0000-9.313946i$     & $0.254023i$              &        & $2$ & $0.0000-11.440951i$    & $0.257438i$\\
       & $3$ &$0.0000-9.567970i$     & $0.253979i$              &        & $3$ & $0.0000-11.698389i$    & $0.257039i$\\
       & $4$ &$0.0000-9.821948i$     & $0.253831i$              &        & $4$ & $0.0000-11.955428i$    & $0.256903i$\\
       & $5$ &$0.0000-10.075779i$    & $0.253767i$              &        & $5$ & $0.0000-12.212331i$    & $0.256617i$\\
       & $6$ &$0.0000-10.329545i$    & $0.253640i$              &        & $6$ & $0.0000-12.468949i$    & $0.256401i$\\
       & $7$ &$0.0000-10.583185i$    & $0.253611i$              &        & $7$ & $0.0000-12.725350i$    & $0.256204i$\\
       [1ex]
\hline\hline 
\end{tabular}
\end{table}

\end{document}